\documentclass[a4paper,11pt]{article}
\usepackage{jheppub}
\hypersetup{hypertexnames=false}
\usepackage[utf8]{inputenc}
\usepackage{amsmath,amssymb}
\usepackage{graphicx}
\usepackage{dcolumn}
\usepackage{bm}
\usepackage{cancel}
\usepackage{makecell}
\usepackage{booktabs}
\usepackage{multirow}
\usepackage{float}
\usepackage{placeins}

\numberwithin{equation}{section}
\newcommand{\D}{d}

\title{Inflationary phase transitions in the early Universe: A Bayesian study with space-based gravitational-wave detectors}

\author[a,b]{Qingyuan Liang}

\author[a,b,c]{Chen Yang}
\emailAdd{yangchen26@ucas.ac.cn}

\author[c,d]{Haipeng An}
\emailAdd{anhp@mail.tsinghua.edu.cn}

\author[a,b]{Huai-Ke Guo}
\emailAdd{guohuaike@ucas.ac.cn}

\affiliation[a]{International Centre for Theoretical Physics Asia-Pacific (ICTP-AP), University of Chinese Academy of Sciences (UCAS), Beijing 100190, China}
\affiliation[b]{Taiji Laboratory for Gravitational Wave Universe (Beijing/Hangzhou), University of Chinese Academy of Sciences (UCAS), Beijing 100190, China}
\affiliation[c]{Department of Physics, Tsinghua University, Beijing 100084, China}
\affiliation[d]{Center for High Energy Physics, Tsinghua University, Beijing 100084, China}

\abstract{
Inflationary phase transitions can generate a stochastic gravitational-wave background that probes primordial physics. We study the detectability and parameter reconstruction of such a signal with a space-based gravitational-wave detector. Using a Taiji-like mission as a benchmark, we construct a realistic data-analysis framework that includes instrumental noise, astrophysical foregrounds and backgrounds, and the $A$, $E$, and $T$ time-delay interferometry channels. The target signal is described in a minimal, model-independent form and analyzed using both Fisher-matrix forecasts and Bayesian inference with nested sampling. We quantify detection significance and parameter-recovery thresholds, showing that, while detection is achievable at moderate signal-to-noise ratios, stronger signals provide more reliable parameter reconstruction. These results offer a realistic assessment of the capability of future space-based missions to probe inflationary phase transitions through stochastic gravitational radiation.
}

\begin{document}
\maketitle

\section{Introduction}\label{sbh}

A stochastic gravitational-wave background (SGWB) of cosmological origin carries fossil information from the early Universe and provides a powerful observational window into physics at energy scales far beyond those accessible to terrestrial experiments~\cite{Caldwell:2022qsj,Roshan:2024qnv,Christensen:2018iqi}. Unlike transient gravitational-wave (GW) signals, a cosmological SGWB is generated by collective processes operating in the primordial plasma and is, therefore, sensitive to the thermal history and symmetry-breaking dynamics of the early Universe~\cite{Caprini:2018mtu,Kosowsky:1992vn,Garcia-Bellido:2007nns}. Recent reviews provide comprehensive discussions of the theoretical foundations and observational prospects of cosmological SGWBs~\cite{LiGong:2024qmt}. Searches for the SGWB span a wide frequency range. In the nanohertz band, pulsar timing array experiments have reported evidence for a stochastic signal~\cite{NANOGrav:2023gor,NANOGrav:2023hde,EPTA:2023sfo,Athron:2023mer}. At higher frequencies, ground-based interferometers including Advanced LIGO, Virgo, and KAGRA operate in the audio band and have placed upper limits in the absence of a detection~\cite{LIGOScientific:2017vtl,Kumar:2024bfe,KAGRA:2020tym,LIGOScientific:2025bgj,LIGOScientific:2025kry}. Space-based interferometers such as Taiji, LISA, and TianQin will probe the millihertz band~\cite{LISA:2017pwj,Robson:2018ifk,LISACosmologyWorkingGroup:2022jok,Hu:2017mde,Ruan:2018tsw,Wu:2018clg,TianQin:2015yph,TianQin:2020hid,Luo:2020bls}.

Among the most promising cosmological mechanisms for generating an SGWB are first-order phase transitions (FOPTs) in the early Universe~\cite{Weir:2017wfa}. Extensive theoretical studies have shown that FOPTs can produce SGWB spectra, sourced by bubble collisions, sound waves, and magnetohydrodynamic turbulence, with amplitudes and characteristic frequencies well suited for detection by space-based missions~\cite{Weir:2017wfa,Mazumdar:2018dfl,Caprini:2015zlo,Caprini:2019egz,Bian:2021ini,Athron:2023xlk,Caprini:2024hue}. Scalar-induced GWs are also an important potential component of the SGWB: Enhanced primordial curvature perturbations source tensor modes at second order when they reenter the horizon~\cite{Ananda:2006af,Baumann:2007zm,Kohri:2018awv,Domenech:2021ztg}. Moreover, inflationary phase transitions (InPTs) can generate a primary GW background together with curvature perturbations, which subsequently source a secondary scalar-induced GW background~\cite{An:2020fff,An:2022cce,An:2023idh,An:2023jxf,Zou:2026wzi,Vagnozzi:2020gtf,Li:2020cjj,An:2024oui,Hu:2025xdt,Sui:2025szm,Bao:2026bgu}. Unlike thermal phase transitions that occur during the radiation-dominated era, InPTs are vacuum phase transitions, in which both the primary GW signal and the curvature perturbations arise from bubble-collision dynamics.

Bayesian reconstruction and component-separation methods for SGWBs have been developed for LISA and related space-based detectors, including generic spectral reconstruction, astrophysical foreground separation, realistic instrumental-noise modeling, phase-transition templates, and Taiji-specific Bayesian studies~\cite{Caprini:2019pxz,Flauger:2020qyi,Kume:2024xvh,Boileau:2022ter,Gowling:2022pzb,Chen:2023zkb,Huang:2025uer}. For scalar-induced backgrounds specifically, recent LISA studies have developed Bayesian reconstruction, parameter-estimation, and model-selection methods for primordial curvature perturbations~\cite{LISACosmologyWorkingGroup:2025vdz,Iovino:2025cdy,Ghaleb:2025xqn}. InPTs can naturally occur in extensions of the standard model, such as grand unified theories~\cite{Hu:2025xdt}, and generate large curvature perturbations without requiring a special inflaton potential. Our work complements this literature by applying detector-level Bayesian parameter estimation to the scalar-induced background generated by an InPT in a Taiji-like detector configuration. 

Although many phenomenological studies of phase-transition-induced SGWBs rely on signal-to-noise ratio (SNR) estimates derived from sensitivity curves, extracting a weak cosmological signal from realistic space-based data remains challenging~\cite{Caprini:2019pxz}. In practice, the SGWB must be disentangled not only from instrumental noise, but also from stochastic astrophysical components, including the confusion foreground produced by unresolved Galactic binaries and the background generated by extragalactic compact binary coalescences~\cite{Boileau:2020rpg,Boileau:2021sni,Caprini:2024hue}. A realistic assessment of detectability, therefore, requires a statistical framework that simultaneously models all relevant components and their correlations~\cite{Biscoveanu:2020gds}.

Space-based interferometers employ time-delay interferometry (TDI) to suppress laser frequency noise, producing multiple approximately noise-orthogonal data combinations, conventionally denoted as the $A$, $E$, and $T$ channels~\cite{Tinto:2014lxa,Smith:2019wny}. Among these, the $T$ channel is largely insensitive to GWs at low frequencies and can, therefore, serve as an internal monitor of instrumental noise~\cite{Adams:2010vc}. Nevertheless, in the absence of cross-correlation between independent detectors, stochastic-background analyses with space-based missions remain more challenging than their ground-based counterparts, where instrumental noise can be efficiently suppressed through interdetector correlations~\cite{Allen:1997ad}.

Recent progress in both theoretical modeling and data-analysis techniques has highlighted the need to combine realistic detector descriptions with statistically rigorous inference methods~\cite{Gowling:2021gcy,Gowling:2022pzb,Boileau:2022ter,Caprini:2024hue,Lewicki:2024xan,Huang:2025uer}. In particular, Bayesian approaches are essential for quantifying parameter uncertainties, evaluating model selection through Bayes factors (BFs), and distinguishing between detection and reliable parameter recovery~\cite{Romano:2016dpx}. Complementary Fisher information matrix (FIM) forecasts provide useful intuition and rapid estimates, but they should be tested against full Bayesian analyses in realistic settings~\cite{OShaughnessy:2013zfw,Porter:2015eha}.

In this work, we develop a comprehensive framework to investigate SGWBs sourced by InPTs with a space-based detector. We construct simulated frequency-domain data for a Taiji-like mission, incorporating instrumental noise, astrophysical foregrounds and backgrounds, and the detector response in the $A$, $E$, and $T$ channels. The InPT contribution is modeled in a minimal, model-independent manner through its characteristic spectral amplitude and reference frequency. We perform parameter inference using both FIM techniques and a full Bayesian analysis based on nested sampling (NS), following our previous studies~\cite{Guan:2025idx,Liang:2025zku}. This enables a direct comparison between forecasted and recovered uncertainties, as well as an explicit evaluation of the BF for signal detection.

Our analysis goes beyond simple detectability estimates by systematically distinguishing among exclusion, detection, and reliable parameter-recovery regimes. We quantify how astrophysical foregrounds and backgrounds affect the reconstruction of the InPT spectrum and determine the signal strength required not only for detection, but also for meaningful inference of the underlying spectral parameters. These results provide a realistic assessment of the scientific reach of future space-based missions in probing InPTs through GWs.

The remainder of this paper is organized as follows. In Sec.~\ref{jvk}, we introduce the theoretical framework for InPTs and the resulting GW spectra. Section~\ref{vdf} describes the Taiji detector response, noise modeling, and the statistical inference pipeline. The results of the parameter-estimation and detectability analyses are presented in Sec.~\ref{ohb}. Finally, Sec.~\ref{qmg} summarizes our findings and discusses their implications for future GW observations.

\section{Gravitational Waves from Inflationary Phase Transitions}\label{jvk}

It is widely accepted that the Universe underwent an epoch of nearly exponential expansion, known as inflation, prior to the standard hot big bang~\cite{Guth:1980zm,Linde:1981mu,Albrecht:1982wi}. The inflationary scenario provides compelling explanations for several cosmological puzzles, including the horizon, flatness, and magnetic monopole problems. Moreover, it generates primordial curvature perturbations that seed the large-scale structure observed today. Observations of the cosmic microwave background and large-scale structure probe the inflationary epoch corresponding to roughly 40--60 e-folds before the end of inflation.

Inflation can be driven by a scalar field $\phi$, called the inflaton. For inflation to end, the inflaton must couple to other spectator fields. Because the inflaton undergoes a large field excursion during inflation, the effective couplings of these spectator fields can evolve with the inflaton~\cite{An:2020fff,An:2022cce}. Here, we consider a single spectator field $\sigma$ with a coupling of the form
\begin{equation}
	V(\phi,\sigma) \supset \frac{1}{2} c \, \phi^{2} \sigma^{2}\ .
\end{equation}
Changes in the effective mass of $\sigma$ may trigger a phase transition, which can be either first order or second order depending on the shape of the $\sigma$ potential. GWs produced during such a phase transition could be observable today and would provide a window into inflationary physics beyond that accessible through cosmic microwave background and large-scale structure observations. In the case of the FOPT during inflation, GWs are generated by bubble collisions, since no plasma is present during that epoch; we refer to these as primary GWs. The characteristic timescale of the transition is determined by the parameter $\beta$,
\begin{equation}
    \beta=\left.-\frac{\D \mathcal{S}_4}{\D t}\right|_{t=t_\star}\ ,
\end{equation}
where $\mathcal{S}_4$ is the four-dimensional Euclidean bounce action associated with quantum tunneling, and $t_\star$ is the time at which the phase transition occurs. For the class of models of interest here, the typical value of $\beta$ is of the order of $\mathcal{O}(10)H_{\rm inf}$, where $H_{\rm inf}$ is the Hubble parameter during inflation. Assuming that the phase transition is instantaneous, the power spectrum of the primary GWs is~\cite{An:2020fff,An:2022cce}
\begin{equation}
\begin{split}
    \Omega_{\rm pri}(f) 
    \simeq  &
    \Omega_{\rm R}
    \left(\frac{H_{\rm inf}}{\beta}\right)^2
    \left(\frac{L}{\rho_{\rm inf}}\right)^2 \left(\cos\left(\frac{f}{f_{\rm ref}}\right)-\frac{\sin(f/f_{\rm ref})}{f/f_{\rm ref}}\right)^2
    \\ &
    \frac{0.3C_{\rm pri}(f/f_{\rm ref})^{-1}}{C_{\rm pri}^4+3(f/f_{\rm ref})^4}\ ,
\end{split}
\end{equation}
where $\Omega_{\rm R}$ is the radiation energy-density fraction today, $L$ is the latent heat released by the phase transition, $\rho_{\rm inf}$ is the total energy density during inflation, and $C_{\rm pri}=1.44\beta/H_{\rm inf}$. The reference frequency $f_{\rm ref}$ is defined as
\begin{equation}
    f_{\rm ref}\simeq 10^{-9}\,{\rm Hz}\ e^{\,40-N_e} \left(\frac{H_{\rm inf}}{10^{14}\,{\rm GeV}}\right)^{1/2}\ ,
\end{equation}
where $N_e$ denotes the number of e-folds between $t_\star$ and the end of inflation. Different phase-transition times $t_\star$, therefore, map to widely separated GW frequency bands, allowing InPT signals to be probed by pulsar timing arrays, space-based detectors, or ground-based interferometers.

At the same time, the backreaction of the phase transition on the inflaton can also produce curvature perturbations, with a power spectrum~\cite{An:2023jxf}
\begin{equation}
    \Delta^2_{\zeta}(f)=A_{\rm ref}\frac{(f/f_{\rm ref})^3}{1+(c_1\,f/f_{\rm ref})^4+(c_2\,f/f_{\rm ref})^9}\ ,
\end{equation}
where $c_1$ and $c_2$ are parameters that depend on $\beta/H_{\rm inf}$, and
\begin{equation}
    A_{\rm ref}=\frac{24}{\epsilon}\left(\frac{M_{\rm Pl}}{\phi_0}\right)^2 \left(\frac{H_{\rm inf}}{\beta}\right)^3 \left(\frac{L}{\rho_{\rm inf}}\right)^2\ ,
\end{equation}
is the reference amplitude, where $\epsilon$ is the slow-roll parameter, $M_{\rm Pl}$ is the reduced Planck mass, and $\phi_0$ is the inflaton field value.

\begin{figure*}[tb!]
    \centering
    \setlength{\fboxsep}{0pt}
    \includegraphics[width=0.98\linewidth]{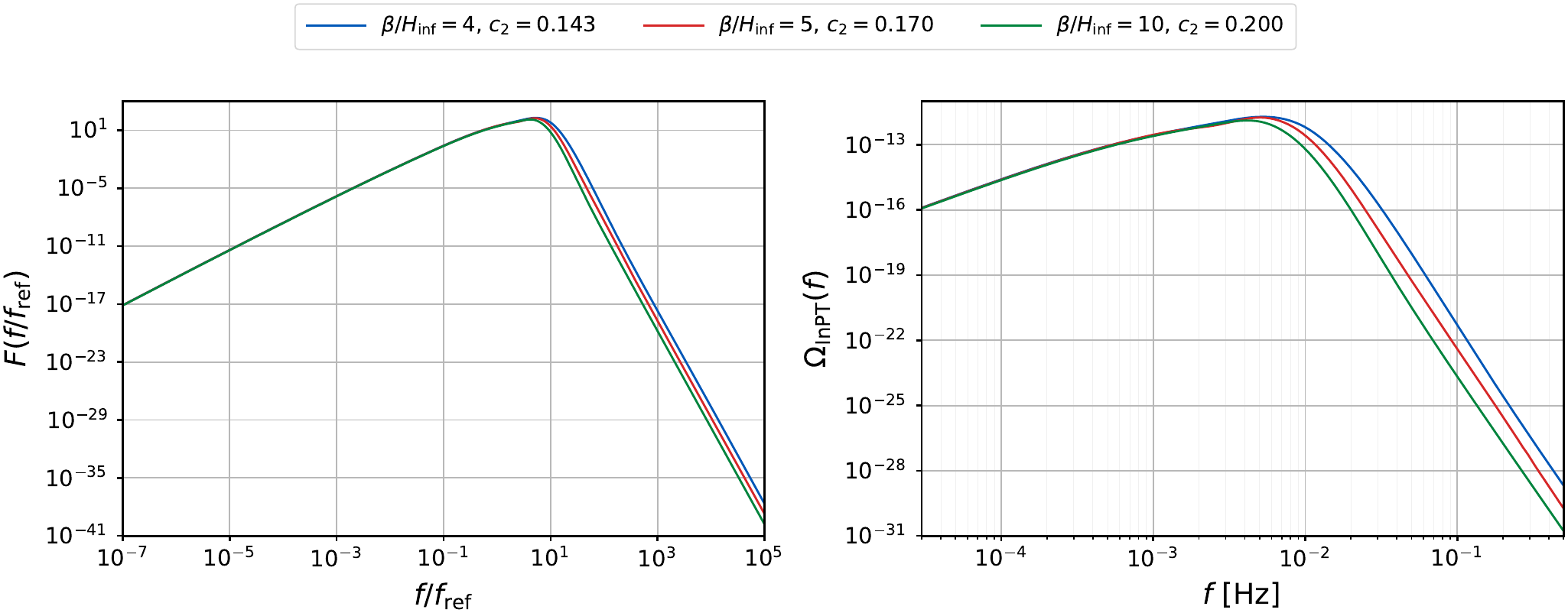}
    \caption{
    Fixed-shape benchmark and template dependence on $\beta/H_{\rm inf}$. The curves compare the shape functions and corresponding normalized InPT spectra for $\beta/H_{\rm inf}=4,5,10$, using $(c_1,c_2)=(0.31,0.143),(0.31,0.170),(0.31,0.200)$, respectively. The red curve is the $\beta/H_{\rm inf}=5$ benchmark used in the main inference analysis. This comparison shows that changing $\beta/H_{\rm inf}$ changes the spectral shape; later PE checks, therefore, compare shape dependence at fixed $(B_{\rm ref},f_{\rm ref})$, rather than a change in peak amplitude.
    }
    \label{fig:beta_template}
\end{figure*}

Since scalar perturbations can source tensor perturbations at second order in the Einstein equations, the curvature perturbations can produce GWs when they reenter the horizon~\cite{Baumann:2007zm,Kohri:2018awv}; we refer to these as secondary GWs. The power spectrum of the secondary GWs is~\cite{An:2023jxf}
\begin{equation}
    \Omega_{\rm sec}(f)=\Omega_{\rm R}\,A_{\rm ref}^2\,F\!\left(\frac{f}{f_{\rm ref}}\right)\ .
\end{equation}
The shape function $F$ is defined as
\begin{equation}
\begin{split}
    F(x)=\int_0^\infty dv\int_{-1}^1 d\mu \left(1-\mu^2\right)^2\frac{v^3}{6}\frac{\bar{I^2}\left(1,v,\sqrt{v^2+1-2v\mu}\right)}{\left(\sqrt{v^2+1-2v\mu}\right)^{3/2}}
    \cdot S(vx)S\left(\sqrt{v^2+1-2v\mu}x\right)\ ,
\end{split}
\end{equation}
where the integration kernel is
\begin{equation}
\begin{split}
    \bar{I^2}\left(k,k_1,k_2\right)=&\frac{1}{2}\left(\frac{3(k_1^2+k_2^2-3k^2)}{4k_1^3k_2^3}\right)^2\left\{\pi^2(k_1^2+k_2^2-3k^2)^2\theta(k_1+k_2-\sqrt{3}k)\right.\\
    &\left.+\left[-4k_1k_2+(k_1^2+k_2^2-3k^2)\log\left|\frac{3k^2-(k_1+k_2)^2}{3k^2-(k_1-k_2)^2}\right|\right]^2\right\}\ ,
\end{split}
\end{equation}

and $S(x)$ is the shape function of the curvature perturbations,
\begin{equation}
    S(x)=\frac{x^3}{1+(c_1 x)^4+(c_2 x)^9}\ .
\end{equation}
The resulting secondary-GW shape function $F(x)$ exhibits a characteristic broken power-law behavior, rising as $F(x)\propto x^3$ for $x\ll1$ and falling steeply as $F(x)\propto x^{-10}$ at high frequencies, with a peak at $x\simeq 5$. The coefficients $c_1$ and $c_2$ determine the shape of $S(x)$ and, therefore, of $F(x)$ for a chosen value of $\beta/H_{\rm inf}$. In the main inference analysis, we use the fixed-shape benchmark $\beta/H_{\rm inf}=5$, for which $c_1=0.31$ and $c_2=0.170$~\cite{An:2023jxf}. We note, however, that other representative choices are also considered in the reference literature; therefore, in this work, we additionally examine two nearby cases, $\beta/H_{\rm inf}=4$ and $10$, to assess the template dependence. The fixed-shape benchmark and these two alternative templates are shown in Fig.~\ref{fig:beta_template}.

For the numerical analysis, $F(x)$ is precomputed using Mathematica's \texttt{NIntegrate} with the \texttt{AdaptiveMonteCarlo} method. We tabulate $F$ on a logarithmic grid $10^{-8}\leq x\leq10^{10}$ with $10^5$ points. In the inference code, $\log_{10}F$ is interpolated as a function of $\log_{10}x$ using a smoothed spline; the JAX (a Python library for automatic differentiation and accelerated numerical computation) likelihood uses a dense tabulated version of the same interpolation for efficient FIM and NS evaluations.

The peak value of the primary GW power spectrum can be parametrized as
\begin{equation}\label{eq:omega_pri_peak}
    \Omega_{\rm GW,\,pri} \propto 7.7\times 10^{-2}\, \Omega_{\rm R} \left(\frac{H_{\rm inf}}{\beta}\right)^5 \left(\frac{L}{\rho_{\rm inf}}\right)^2 ,
\end{equation}
while the peak energy density of the secondary component scales as
\begin{equation}\label{eq:omega_sec_peak}
    \Omega_{\rm GW,\,sec}\propto 1.1\times 10^{5}\, \Omega_{\rm R}\frac{1}{\epsilon^2} \left(\frac{M_{\rm Pl}}{\phi_0}\right)^4 \left(\frac{H_{\rm inf}}{\beta}\right)^6 \left(\frac{L}{\rho_{\rm inf}}\right)^4 .
\end{equation}
The secondary GW component is parametrically larger than the primary one. Because the quantitative validity check is most naturally interpreted after the Taiji detectability contours have been introduced, we collect the primary--secondary comparison in Appendix~\ref{app:primary_secondary}. The comparison shows that the breakdown boundary of the secondary-only approximation lies far below the Taiji-accessible region. In the remainder of this work, we therefore focus exclusively on the secondary GW contribution, which dominates the observable signal over the relevant parameter space. The present-day GW energy-density spectrum can be written as
\begin{equation}
    \Omega_{\rm InPT}(f) \simeq \Omega_{\rm GW,\,sec}(f)= \Omega_{\rm R}\, A_{\rm ref}^2\, F\!\left(\frac{f}{f_{\rm ref}}\right)\ .
\label{tgv}
\end{equation}

\section{Data-Analysis Framework with Taiji} \label{vdf}

Taiji is a planned space-based GW observatory operating in the millihertz band, in which stochastic signals from primordial processes such as InPTs may be detectable. The mission consists of three drag-free spacecraft forming a triangular constellation in heliocentric orbit, and laser phase measurements along the three arms are processed using TDI~\cite{Tinto:2004wu} to suppress laser frequency noise. This procedure yields three orthogonal data channels, conventionally denoted by $A$, $E$, and $T$. In the frequency domain, the one-sided power spectral density of each channel can be written as~\cite{Guan:2025idx}
\begin{equation}
    P_a(f) = S_a(f) + N_a(f),
    \qquad a \in \{A,E,T\},
\end{equation}
where $S_a(f)$ represents the GW-induced contribution and $N_a(f)$ denotes the instrumental noise.

For an isotropic SGWB, the signal contribution is related to the GW energy-density spectrum through~\cite{Guan:2025idx}
\begin{equation}
    S_a(f)
    =
    \frac{3 H_0^2}{4 \pi^2}
    \frac{\Omega_{\rm GW}(f)}{f^3}
    \mathcal{R}_a(f).
\end{equation}
Here, $\mathcal{R}_a(f)$ denotes the response function of the corresponding TDI channel. Throughout this work, we adopt the standard Taiji response functions and assume equal, time-independent arm lengths.

The instrumental noise of Taiji is modeled as the combination of acceleration noise and optical metrology noise, whose amplitude spectral densities are given by~\cite{Guan:2025idx}
\begin{equation}\label{efv}
    \begin{aligned}
        \sqrt{S_{\rm acc}(f)} &=
        N_{\rm acc}
        \sqrt{1 + \left( \frac{0.4~{\rm mHz}}{f} \right)^2}
        \sqrt{1 + \left( \frac{f}{8~{\rm mHz}} \right)^4}
        \left( \frac{{\rm m}}{{\rm s}^2 \sqrt{\rm Hz}} \right), \\
        \sqrt{S_{\rm oms}(f)} &=
        \delta x
        \sqrt{1 + \left( \frac{2~{\rm mHz}}{f} \right)^4}
        \left( \frac{{\rm m}}{\sqrt{\rm Hz}} \right),
    \end{aligned}
\end{equation}
where we adopt the nominal values $N_{\rm acc} = 3 \times 10^{-15}$ and $\delta x = 8 \times 10^{-12}$ for Taiji. These noise sources combine into the power spectral densities of the $A$, $E$, and $T$ channels as
\begin{equation}
    N_A(f) = N_E(f) = N_1(f) - N_2(f),
    \qquad
    N_T(f) = N_1(f) + 2 N_2(f),
\end{equation}
with
\begin{align}
    N_1(f) &=
    \frac{1}{L^2}
    \left[
        4 S_{\rm oms}(f)
        + \frac{8}{(2\pi f)^4}
        \left( 1 + \cos^2 \frac{f}{f_*} \right)
        S_{\rm acc}(f)
    \right]
    |W(f)|^2, \\
    N_2(f) &=
    -\frac{1}{L^2}
    \left[
        2 S_{\rm oms}(f)
        + \frac{8}{(2\pi f)^4}
        S_{\rm acc}(f)
    \right]
    \cos \frac{f}{f_*}
    |W(f)|^2 .
\end{align}
Here, $L$ denotes the arm length of Taiji, $f_* = c/(2\pi L)$, and $W(f) = 1 - e^{-2 i f / f_*}$ is the TDI transfer function.

In realistic observations, the cosmological SGWB is accompanied by astrophysical foregrounds and backgrounds. The total GW energy-density spectrum is modeled as
\begin{equation}
    \Omega_{\rm GW}(f)
    =
    \Omega_{\rm DWD}(f)
    +
    \Omega_{\rm PL}(f)
    +
    \Omega_{\rm InPT}(f).
\end{equation}
The Galactic foreground from unresolved double white-dwarf binaries is described by a broken power-law form~\cite{Chen:2023zkb,Boileau:2022ter}:
\begin{equation}\label{wbv}
    \Omega_{\rm DWD}(f)
    =
    \frac{
        A_1 \left( f / f_* \right)^{\alpha_1}
    }{
        1 + A_2 \left( f / f_* \right)^{\alpha_2}
    }.
\end{equation}
The extragalactic astrophysical background is modeled as a single power law~\cite{Boileau:2020rpg}:
\begin{equation}\label{emv}
    \Omega_{\rm PL}(f)
    =
    \Omega_{\rm ast}
    \left( \frac{f}{10^{-3} \, {\rm Hz}} \right)^{\varepsilon},
\end{equation}
where $\Omega_{\rm ast}$ is defined at the reference frequency $10^{-3}\,{\rm Hz}$ and $\varepsilon$ is the spectral index.

The SGWB sourced by an InPT can be parametrized according to Eq.~\eqref{tgv} as
\begin{equation}\label{gbc}
    \Omega_{\rm InPT}(f)
    =
    \Omega_{\rm R} A_{\rm ref}^2\,
    F\!\left( \frac{f}{f_{\rm ref}} \right)
    \equiv
    B_{\rm ref}\,
    F\!\left( \frac{f}{f_{\rm ref}} \right)\ .
\end{equation}
The dimensionless function $F(f/f_{\rm ref})$ encodes the spectral shape of the InPT signal. Here, $F$ is the fixed InPT shape function defined in Eqs.~(2.8)--(2.10), evaluated for the benchmark choice $\beta/H_{\rm inf}=5$. The quantities $f_{\rm ref}$ and $A_{\rm ref}$ are the same quantities defined in Sec.~\ref{jvk}.

For convenience, we combine $\Omega_{\rm R}A_{\rm ref}^2$ into the effective amplitude $B_{\rm ref}\equiv\Omega_{\rm R}A_{\rm ref}^2$. With this reparametrization, the cosmological contribution from an InPT is fully characterized by two parameters, $(B_{\rm ref},\, f_{\rm ref})$, which are taken to be the only free parameters describing the inflationary signal in our analysis.

As described in our previous study~\cite{Guan:2025idx}, the total observation time is divided into \(N_0 = 126\) statistically independent segments of duration \(T = 10^6\,\mathrm{s}\), resulting in an effective observing time \(T_t = N_0 T\) of about 4 yr. The analysis is carried out in the frequency domain with a resolution \(\Delta f = 1/T\), assuming stationary Gaussian noise within each segment. The likelihood is constructed from the Fourier-domain data of the \(A\), \(E\), and \(T\) channels, with different frequency bins assumed to be statistically independent. Parameter inference is performed within a Bayesian framework, while the FIM is used for rapid forecasting and consistency checks. We employ NS to sample the full posterior distributions of the model parameters and to compute the Bayesian evidence required for BF evaluation. Before defining the SNRs, we specify the notation for the spectral components. In a quantity such as $S_A^X(f)$, the subscript $A$ denotes the TDI $A$ channel, while the superscript $X$ labels the physical signal component contributing to that channel spectrum; it is a component label rather than an exponent. We use $X={\rm InPT}$ for the inflationary phase-transition signal, $X={\rm DWD}$ for the unresolved Galactic double-white-dwarf foreground, and $X={\rm ast}$ for the extragalactic astrophysical SGWB. The instrumental noise contribution is denoted separately by $N_A(f)$. With this convention, the absolute and relative SNRs used below are defined by

\begin{align}
\mathrm{SNR}_a^2
&=
2T_t\int_{f_{\min}}^{f_{\max}}
\left[
\frac{S_A^{\rm InPT}(f)}{N_A(f)}
\right]^2\mathrm{d} f,\\
\mathrm{SNR}_r^2
&=
2T_t\int_{f_{\min}}^{f_{\max}}
\left[
\frac{S_A^{\rm InPT}(f)}
{N_A(f)+S_A^{\rm DWD}(f)+S_A^{\rm ast}(f)}
\right]^2\mathrm{d} f,
\end{align}
where
\begin{equation}
S_A^{\rm InPT}(f)
=
S_A^{\rm DWD+ast+InPT}(f)
-
S_A^{\rm DWD+ast}(f).
\end{equation}
Here, the compound superscript, for example, ${\rm DWD+ast+InPT}$, denotes the sum of the corresponding stochastic signal components in the $A$ channel, not including the instrumental noise. Thus, subtracting the spectrum without the InPT contribution from the spectrum with it isolates $S_A^{\rm InPT}(f)$. Here $f_{\min}=3\times10^{-5}\,{\rm Hz}$, $f_{\max}=0.5\,{\rm Hz}$, and $T_t=126\times10^6\,{\rm s}$. Thus, $\mathrm{SNR}_a$ measures the InPT signal against instrumental noise, whereas $\mathrm{SNR}_r$ includes the DWD foreground and astrophysical stochastic-background components in the denominator.

\section{Results}\label{ohb}

In this section, we carry out the data-analysis procedure introduced in Sec.~\ref{vdf}. We perform Bayesian inference using NS as implemented in \texttt{Bilby.dynesty}, considering a total of ten parameters: the instrumental-noise parameters $N_\mathrm{acc}$ and $\delta x$ [see Eq.~\eqref{efv}], the astrophysical-foreground parameters $A_1$, $\alpha_1$, $A_2$, and $\alpha_2$ [see Eq.~\eqref{wbv}], the astrophysical-background parameters $\Omega_{\rm ast}$ and $\varepsilon$ [see Eq.~\eqref{emv}], and the InPT parameters $B_{\rm ref}$ and $f_{\rm ref}$ [see Eq.~\eqref{gbc}]. The prior ranges adopted for these parameters are summarized in Table~\ref{sbp} and are used consistently throughout the NS analysis.

\begin{figure*}[tb!]
    \centering
    \includegraphics[width=0.9\linewidth]{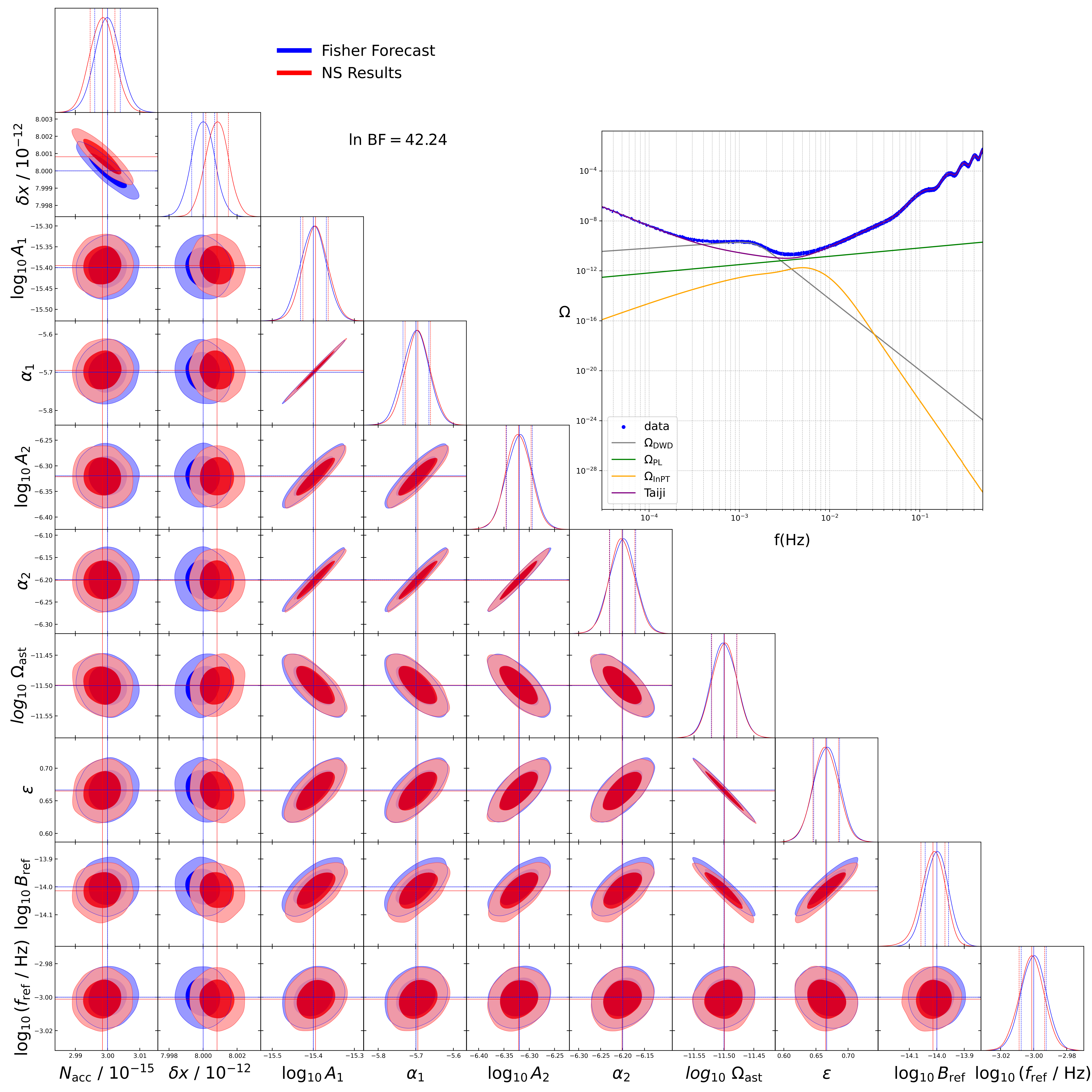}
    \caption{
    Corner plot of the posterior distributions for the ten model parameters obtained from NS (red), compared with the corresponding FIM confidence ellipses (blue). Dark and light shaded regions indicate the $68\%$ and $95\%$ credible levels, respectively. Red crosshairs denote the posterior means inferred from NS, while blue crosshairs mark the injected fiducial values. The diagonal panels show the marginalized one-dimensional distributions, with dashed lines indicating the $1\sigma$ intervals (red for NS and blue for FIM). The inset displays the injected stochastic spectra, including the astrophysical foreground (gray), the astrophysical background (green), and the InPT signal (orange), together with the Taiji noise curve (purple). For this injection, the resulting SNRs are $\mathrm{SNR}_a=118$ and $\mathrm{SNR}_r=66$, and the NS analysis yields $\ln \mathrm{BF}=42.24$ when comparing models with and without an InPT component. Overall agreement between the NS and FIM results is observed, with small discrepancies arising from non-Gaussian posteriors and statistical fluctuations.
    }
    \label{egl}
\end{figure*}

\begin{table}[tb!]
\centering
    \begin{tabular}{lc}
    \hline\hline
    Parameter & Prior range (uniform) \\
    \hline
    $N_{\rm acc} / 10^{-15}$                 & $(0,\,20)$      \\
    $\delta x / 10^{-12}$                   & $(0,\,20)$      \\
    $\log_{10} A_1$              & $(-17,\,-13)$   \\
    $\alpha_1$                   & $(-10,\,-3)$    \\
    $\log_{10} A_2$              & $(-10,\,-2)$    \\
    $\alpha_2$                   & $(-10,\,-1)$    \\
    $\log_{10} \Omega_{\rm ast}$ & $(-15,\,-8)$    \\
    $\varepsilon$                & $(-2,\,3)$      \\
    $\log_{10} B_{\rm ref}$      & $(-16,\,-9)$    \\
    $\log_{10} (f_{\rm ref}/\rm Hz)$      & $(-5,\,-1)$     \\
    \hline\hline
    \end{tabular}
\caption{Prior ranges adopted for the instrumental-noise, astrophysical-foreground and -background, and InPT parameters in the NS analysis.}
\label{sbp}
\end{table}

After performing NS, we obtain the full posterior distributions of the model parameters, as shown in Fig.~\ref{egl}. The injected and recovered parameter values are summarized in Table~\ref{sok}. In the corner plot, the red contours indicate the posterior distributions obtained from NS, while the blue contours represent the corresponding confidence ellipses predicted by the FIM. The two sets of results show good overall agreement, indicating that in this regime the FIM provides an efficient and reliable approximation for forecasting parameter uncertainties.

Nevertheless, small deviations between the NS and FIM results are visible, which can be attributed to mild non-Gaussian features in the posterior distributions and statistical fluctuations arising from the data-generation process. In Fig.~\ref{egl}, the dark and light shaded regions correspond to the $1\sigma$ and $2\sigma$ confidence regions, respectively. The red crosshairs mark the posterior means recovered from NS, while the blue crosshairs indicate the injected fiducial values. The marginalized one-dimensional distributions are shown along the diagonal, where the dashed lines denote the $1\sigma$ intervals (red for NS and blue for FIM).

\begin{figure*}[tb!]
    \centering
    \includegraphics[height=0.25\linewidth]{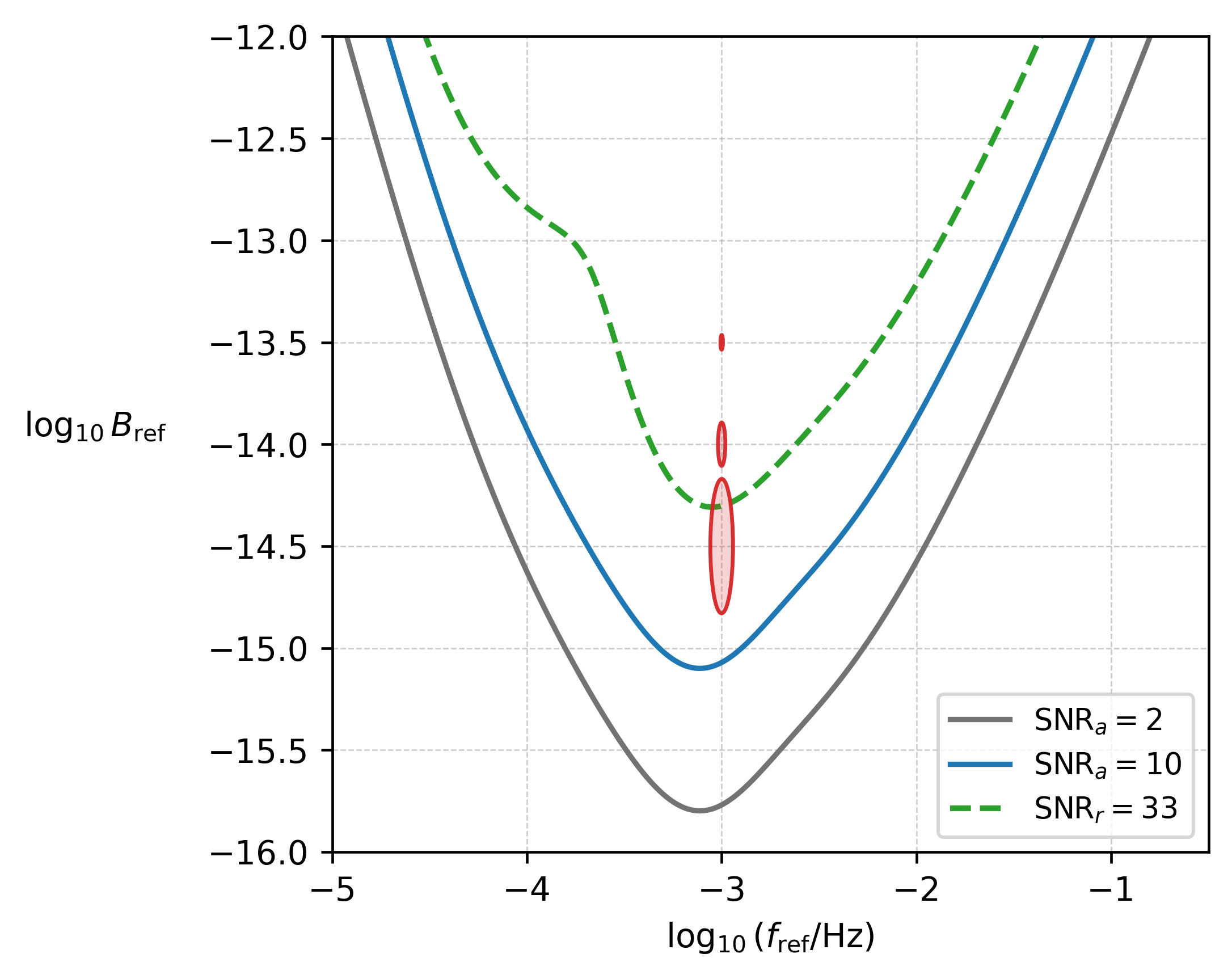}\hfill
    \includegraphics[height=0.25\linewidth]{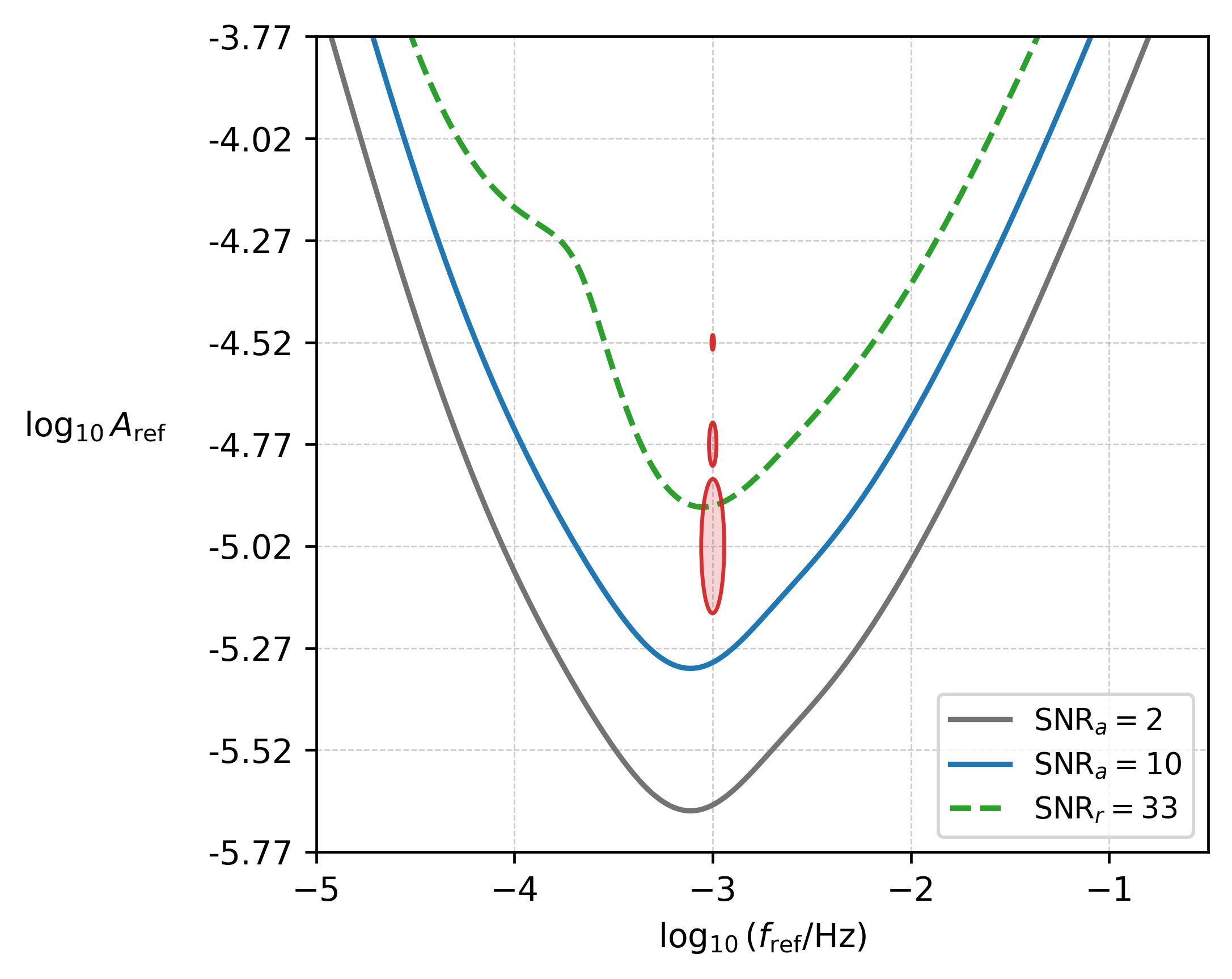}\hfill
    \includegraphics[height=0.25\linewidth]{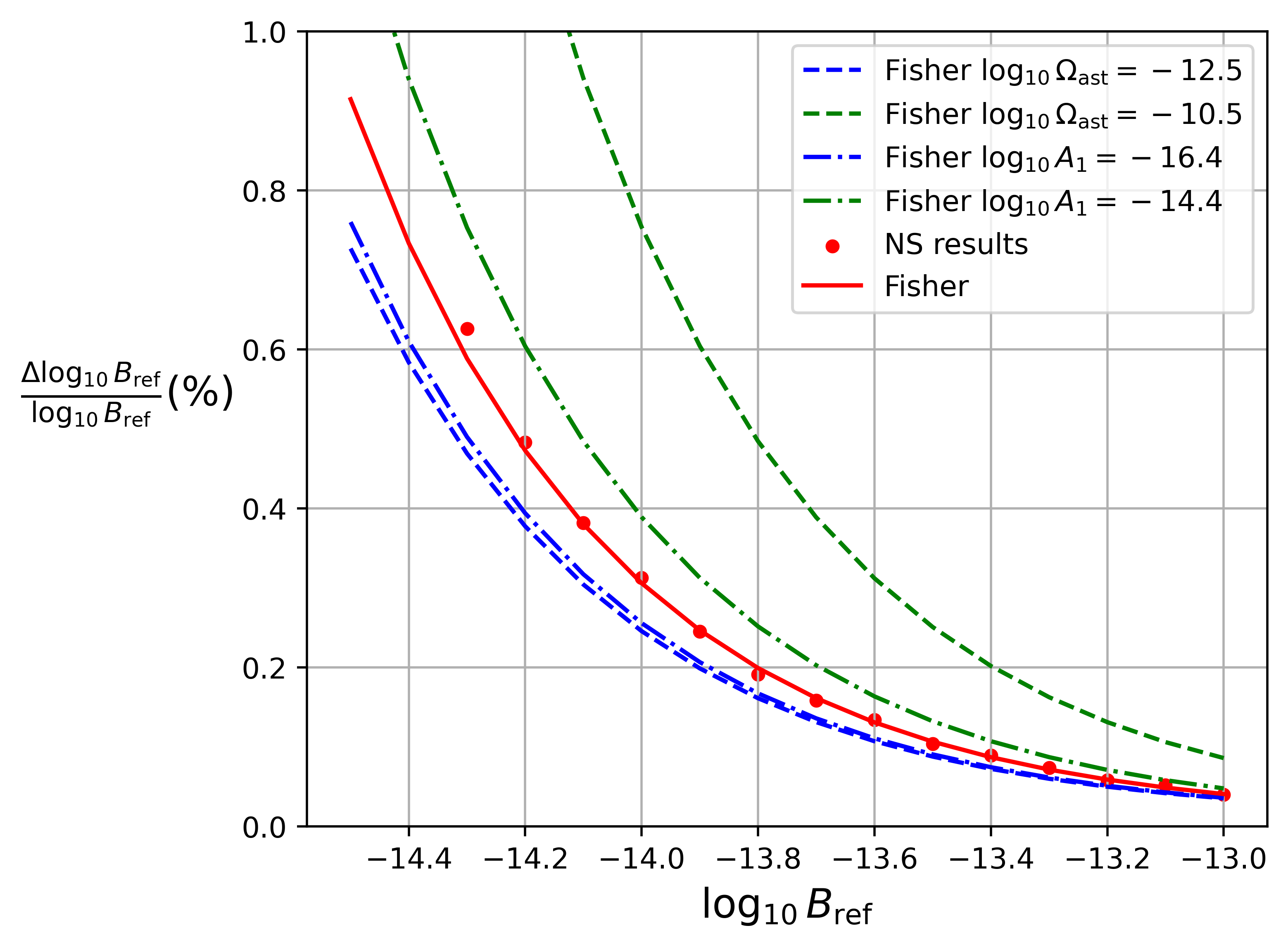}
    \caption{
    Detectability and parameter recovery of the InPT signal. The left panel shows exclusion and detection contours in the $(\log_{10}(f_{\rm ref}/{\rm Hz}),\,\log_{10} B_{\rm ref})$ plane. The gray and blue curves correspond to absolute SNRs $\mathrm{SNR}_{a}=2$ and $\mathrm{SNR}_{a}=10$, respectively, while the green dashed curve marks the relative-SNR recovery benchmark $\mathrm{SNR}_{r}=33$. Red ellipses denote representative $2\sigma$ confidence regions predicted by the FIM at $\log_{10}(f_{\rm ref}/{\rm Hz})=-3$ and $\log_{10} B_{\rm ref}=-14.5,-14,-13.5$. The middle panel shows the same contours expressed in terms of $\log_{10} A_{\rm ref}$, obtained from $B_{\rm ref}$ via Eq.~\eqref{gbc}. The right panel presents the relative uncertainty in $\log_{10} B_{\rm ref}$ as a function of its injected value for different astrophysical foreground and background amplitudes. The solid red curve corresponds to the fiducial model, red markers indicate NS results at selected injection points, and the dashed (dot-dashed) curves represent enhanced or suppressed astrophysical background (foreground) contributions.
    }
    \label{elv}
\end{figure*}

\begin{table*}[t!]
\centering
\small
\renewcommand{\arraystretch}{0.88}
\begin{tabular}{lcccc}
\hline\hline
Parameter & Fiducial value & NS recovery & Fisher unc.\ (\%) & NS unc.\ (\%) \\
\hline
$N_{\rm acc} / 10^{-15}$ 
& $3.000$ & $2.998$ & $0.132$ & $0.129$ \\
$\delta x / 10^{-12}$ 
& $8.000$ & $8.001$ & $0.008$ & $0.008$ \\
$\log_{10} A_1$ 
& $-15.400$ & $-15.395$ & $0.204$ & $0.200$ \\
$\alpha_1$ 
& $-5.700$ & $-5.695$ & $0.602$ & $0.593$ \\
$\log_{10} A_2$ 
& $-6.320$ & $-6.321$ & $0.401$ & $0.393$ \\
$\alpha_2$ 
& $-6.200$ & $-6.201$ & $0.475$ & $0.467$ \\
$\log_{10} \Omega_{\rm ast}$ 
& $-11.500$ & $-11.499$ & $0.187$ & $0.183$ \\
$\varepsilon$ 
& $0.667$ & $0.665$ & $3.050$ & $2.989$ \\
$\log_{10} B_{\rm ref}$ 
& $-14.000$ & $-14.014$ & $0.306$ & $0.312$ \\
$\log_{10} (f_{\rm ref} / \rm Hz)$
& $-3.000$ & $-3.001$ & $0.253$ & $0.256$ \\
\hline\hline
\end{tabular}
\caption{Injected parameter values, posterior means inferred from NS, and relative uncertainties obtained from the FIM and NS analyses.}
\label{sok}
\vspace{0.8em}
\begin{tabular}{cccccc}
\hline\hline
$\log_{10} B_{\rm ref}$ 
& FIM unc. (\%) 
& NS unc. (\%) 
& $\mathrm{SNR}_a$
& $\mathrm{SNR}_r$ 
& $\ln \mathrm{BF}$ \\
\hline
$-14.3$ & $0.588$ & $0.626$ & $58.993$ & $33.177$ & $7.517$  \\
$-14.2$ & $0.472$ & $0.483$ & $74.268$ & $41.768$ & $14.673$ \\
$-14.1$ & $0.380$ & $0.382$ & $93.498$ & $52.583$ & $26.307$ \\
$-14.0$ & $0.306$ & $0.312$ & $117.707$ & $66.197$ & $42.236$ \\
$-13.9$ & $0.246$ & $0.245$ & $148.185$ & $83.338$ & $73.518$ \\
$-13.8$ & $0.199$ & $0.191$ & $186.554$ & $104.916$ & $128.998$ \\
$-13.7$ & $0.161$ & $0.158$ & $234.857$ & $132.081$ & $188.963$ \\
$-13.6$ & $0.131$ & $0.134$ & $295.668$ & $166.281$ & $279.363$ \\
$-13.5$ & $0.106$ & $0.103$ & $372.224$ & $209.335$ & $509.630$ \\
$-13.4$ & $0.087$ & $0.089$ & $468.602$ & $263.537$ & $746.782$ \\
$-13.3$ & $0.071$ & $0.074$ & $589.935$ & $331.773$ & $1148.164$ \\
$-13.2$ & $0.059$ & $0.058$ & $742.684$ & $417.678$ & $2008.068$ \\
$-13.1$ & $0.049$ & $0.052$ & $934.984$ & $525.825$ & $2999.619$ \\
$-13.0$ & $0.040$ & $0.040$ & $1177.075$ & $661.975$ & $4987.875$ \\
\hline\hline
\end{tabular}
\caption{
Relative uncertainties in $\log_{10} B_{\rm ref}$ obtained from FIM forecasts and NS, together with the corresponding absolute and relative SNRs, $\mathrm{SNR}_a$ and $\mathrm{SNR}_r$, and BFs for different injected InPT amplitudes.
}
\label{ebl}
\end{table*}
We next assess the detectability of the InPT contribution by constructing exclusion and detection contours in the $(\log_{10} (f_{\rm ref} / \rm Hz),\,\log_{10} B_{\rm ref})$ plane, as shown in Fig.~\ref{elv}. Following standard conventions in the literature, we adopt an absolute SNR of 10 as the detection threshold~\cite{Boileau:2025jkv}, while an SNR of 2 defines the exclusion limit~\cite{Chen:2024ikn}, above which a signal can be excluded in the absence of a detection. These criteria are illustrated in Fig.~\ref{elv} by the blue (detection) and gray (exclusion) contours, respectively, in all panels except the rightmost one.

While the detection threshold specifies the minimum signal strength required to claim a detection, it does not guarantee accurate parameter inference. For this reason, the exclusion and detection curves in Fig.~\ref{elv} are drawn with the absolute SNR, $\mathrm{SNR}_{a}=2$ and $\mathrm{SNR}_{a}=10$, whereas the green dashed curve uses the relative SNR, $\mathrm{SNR}_{r}=33$, as the benchmark for accurate parameter recovery in the presence of astrophysical foregrounds and backgrounds. In our representative NS sequence, this benchmark corresponds to the onset of strong evidence and stable recovery; for example, at $\log_{10}B_{\rm ref}=-14.3$ and $\log_{10}(f_{\rm ref}/{\rm Hz})=-3$, the analysis gives $\mathrm{SNR}_{r}=33.177$ and $\ln\mathrm{BF}=7.517$.

The red ellipses in Fig.~\ref{elv} indicate representative $2\sigma$ confidence regions predicted by the FIM at selected test points with $\log_{10} (f_{\rm ref} / \rm Hz)=-3$ and $\log_{10} B_{\rm ref}=-14.5,\,-14,\,-13.5$. As expected, increasing $B_{\rm ref}$ leads to progressively smaller ellipses, reflecting improved constraints for stronger InPT signals. The left and middle panels in Fig.~\ref{elv} display the detection and exclusion contours in terms of $B_{\rm ref}$ (left) and the equivalent amplitude parameter $A_{\rm ref}$ (middle), related through Eq.~\eqref{gbc}.

The inset panel in the upper-right corner displays the individual spectral components: the astrophysical foreground (gray), the astrophysical background (green), and the InPT signal (orange), together with the Taiji noise curve (purple). For this injection, the absolute and relative SNRs are $\mathrm{SNR}_a = 118$ and $\mathrm{SNR}_r = 66$, respectively, and the NS analysis yields $\ln \mathrm{BF} \simeq 42.24$ when comparing models with and without an InPT component. The relative uncertainties obtained from both NS and FIM are listed in Table~\ref{sok}. According to the Cram\'er--Rao lower bound, the FIM uncertainties are expected to be comparable to or smaller than those inferred from NS; the small discrepancies observed are consistent with the statistical fluctuations described above.

\begin{figure*}[tb!]
    \centering
    \setlength{\fboxsep}{0pt}
    \includegraphics[width=0.88\linewidth]{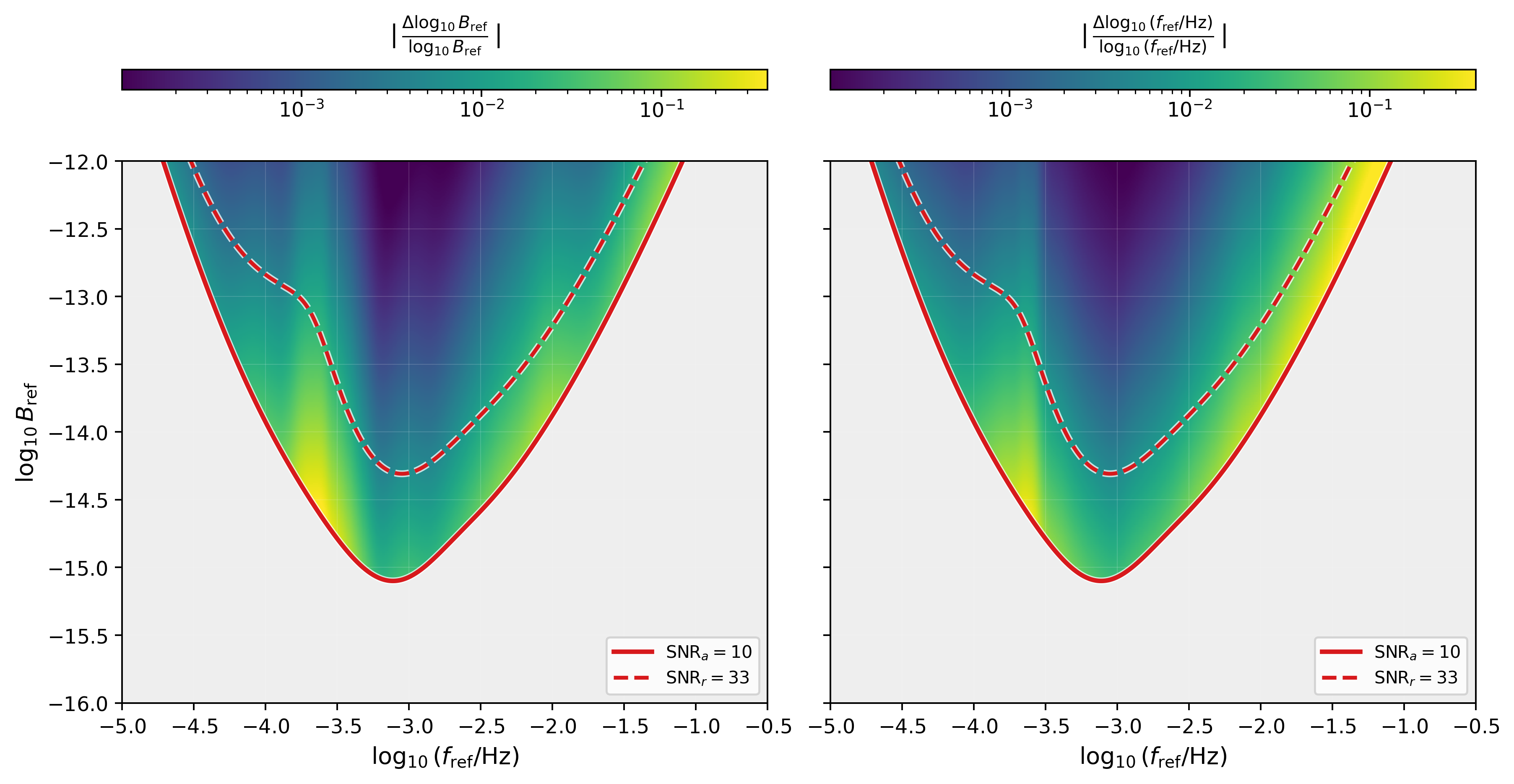}
    \caption{
    Full-plane FIM reconstruction-uncertainty maps in the $\left(\log_{10}(f_{\rm ref}/{\rm Hz}),\log_{10}B_{\rm ref}\right)$ plane. Since the inference variables are logarithmic, the plotted quantities are the relative uncertainties of the reconstructed logarithmic parameters, $\lvert \frac{\Delta\log_{10}B_{\rm ref}}{\log_{10}B_{\rm ref}} \rvert$ and $\lvert \frac{\Delta\log_{10}(f_{\rm ref}/{\rm Hz})}{\log_{10}(f_{\rm ref}/{\rm Hz})} \rvert$. The solid contour marks the absolute-SNR detection threshold $\mathrm{SNR}_{a}=10$, while the dashed contour marks the relative-SNR recovery benchmark $\mathrm{SNR}_{r}=33$; lower values indicate where a detected InPT signal can also be accurately characterized.
    }
    \label{fig:reconstruction_heatmaps}
\end{figure*}

The full-plane FIM maps in Fig.~\ref{fig:reconstruction_heatmaps} complement the one-dimensional uncertainty slice shown in the right panel in Fig.~\ref{elv} by showing where a detected InPT signal can also be quantitatively characterized. The reconstruction precision is not uniform across the Taiji-sensitive region. It is best in the central millihertz band and for larger $B_{\rm ref}$, where the InPT spectrum has strong overlap with the most sensitive part of the detector response. The uncertainties grow toward the low- and high-frequency edges and close to the $\mathrm{SNR}_a=10$ detection boundary, while the $\mathrm{SNR}_r=33$ contour marks the approximate onset of reliable parameter recovery. Thus, the detection contour should be interpreted only as the boundary for identifying a signal, while the lower-uncertainty regions in Fig.~\ref{fig:reconstruction_heatmaps} indicate where both the signal amplitude and the transition frequency can be recovered well enough for physical interpretation.

Additional PE checks for a lower-SNR injection and for alternative fixed-shape templates are collected in Appendix~\ref{app:supplementary_pe}, since they support the robustness interpretation but are not needed for the main detectability discussion.

Finally, we examine how astrophysical foregrounds and backgrounds affect the recoverability of the InPT signal. The right panel in Fig.~\ref{elv} shows the relative uncertainty in $\log_{10} B_{\rm ref}$ as a function of its injected value under different assumptions for the strengths of the astrophysical components. The solid red curve corresponds to the fiducial injection listed in Table~\ref{sok}, with only $B_{\rm ref}$ varied, while the red markers denote results obtained directly from NS at selected injection points. The dashed and dot-dashed curves represent cases with enhanced or suppressed astrophysical background and foreground amplitudes, respectively. Stronger astrophysical contributions degrade the precision of the cosmological parameter recovery, whereas weaker ones lead to tighter constraints. For the NS results, the corresponding recovered SNRs and BFs are reported in Table~\ref{ebl}.

Taken together, these results have direct implications for the study of InPTs. In our parametrization, the Bayesian recovery of $(B_{\rm ref}, f_{\rm ref})$ can be translated into constraints on the characteristic energy release, timescale, and occurrence epoch of the transition during inflation. Here, the energy-release variable is $L/\rho_{\rm inf}$, namely, the released energy density normalized by the inflationary background energy density; the timescale is the inverse transition duration $\beta^{-1}$, usually expressed through $H_{\rm inf}/\beta$ or $\beta/H_{\rm inf}$; and the occurrence epoch is the number of e-folds $N_e$ before the end of inflation, equivalently encoded by $f_{\rm ref}$ through Eq.~(2.4). In particular, $f_{\rm ref}$ encodes when the phase transition takes place relative to the end of inflation, while $B_{\rm ref}$ is related to the transition strength through the combination of parameters entering the secondary GW amplitude. Since
\begin{equation}
B_{\rm ref}=\Omega_{\rm R}A_{\rm ref}^2,\qquad
A_{\rm ref}=\frac{24}{\epsilon}
\left(\frac{M_{\rm Pl}}{\phi_0}\right)^2
\left(\frac{H_{\rm inf}}{\beta}\right)^3
\left(\frac{L}{\rho_{\rm inf}}\right)^2,
\end{equation}
a measurement of $B_{\rm ref}$ does not by itself determine a single microscopic quantity. As a concrete example, fixing $M_{\rm Pl}/\phi_0=1$ and $H_{\rm inf}/\beta=1/5$ gives
\begin{equation}
A_{\rm ref}=\sqrt{\frac{B_{\rm ref}}{\Omega_{\rm R}}}
=
\frac{0.192}{\epsilon}
\left(\frac{L}{\rho_{\rm inf}}\right)^2,
\end{equation}

\begin{figure}[t!]
    \centering
    \setlength{\fboxsep}{0pt}
    \includegraphics[width=0.80\linewidth]{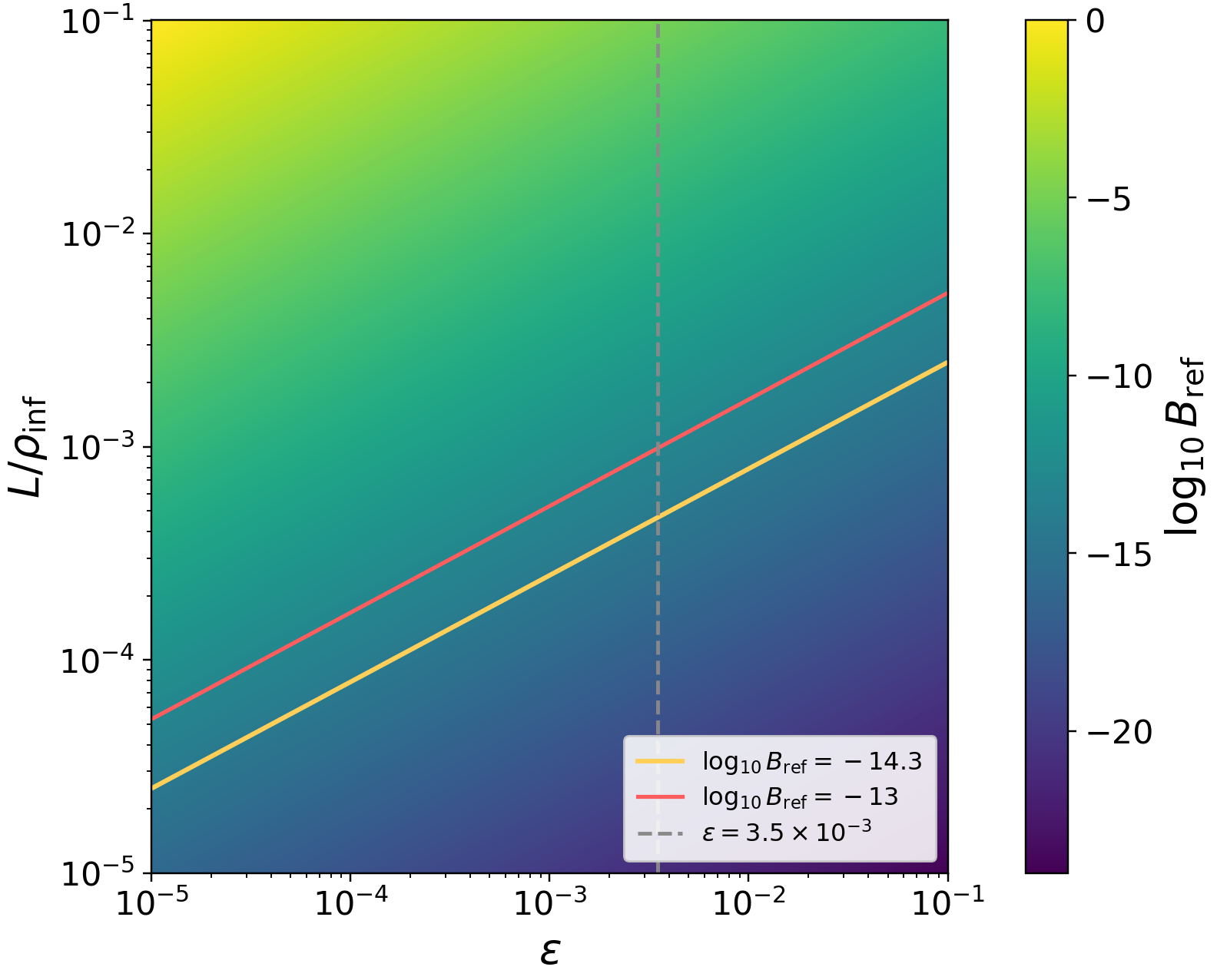}
    \vspace{-0.5em}
    \caption{
    Worked physical interpretation of $B_{\rm ref}$ for $M_{\rm Pl}/\phi_0=1$ and $H_{\rm inf}/\beta=1/5$. The heat map shows $\log_{10}B_{\rm ref}$ in the $(\epsilon,L/\rho_{\rm inf})$ plane, where $\epsilon$ is the slow-roll parameter entering the InPT amplitude in Eq.~(4.1). The gray dashed vertical line marks the benchmark value $\epsilon\simeq3.5\times10^{-3}$.
    }
    \label{fig:physical_scan}
    \vspace{-1.2em}
\end{figure}

Thus, the conversion from $B_{\rm ref}$ back to $L/\rho_{\rm inf}$ uses the radiation-fraction normalization adopted in the plotted spectra, $\Omega_{\rm R}\simeq3.5\times10^{-5}$. Figure~\ref{fig:physical_scan} shows the resulting scan in the $(\epsilon,L/\rho_{\rm inf})$ plane, where $\epsilon$ denotes the slow-roll parameter entering the InPT amplitude above. The two $B_{\rm ref}$ contours, $\log_{10}B_{\rm ref}=-14.3$ and $-13.0$, mark the recovery range in Table~\ref{ebl}. To provide a familiar orientation scale, we also include one benchmark with $\epsilon\simeq3.5\times10^{-3}$ in the plot. In this benchmark, $L/\rho_{\rm inf}$ is around $3\times10^{-4}$.

The transition epoch associated with a measured $f_{\rm ref}$ is also conditional on the inflationary scale. From Eq.~(2.4),
\begin{equation}
N_e\simeq
40-\ln\left(\frac{f_{\rm ref}}{10^{-9}{\rm Hz}}\right)
+
\frac{1}{2}\ln\left(\frac{H_{\rm inf}}{10^{14}{\rm GeV}}\right).
\end{equation}

\begin{figure}[t!]
    \centering
    \setlength{\fboxsep}{0pt}
    \includegraphics[width=0.80\linewidth]{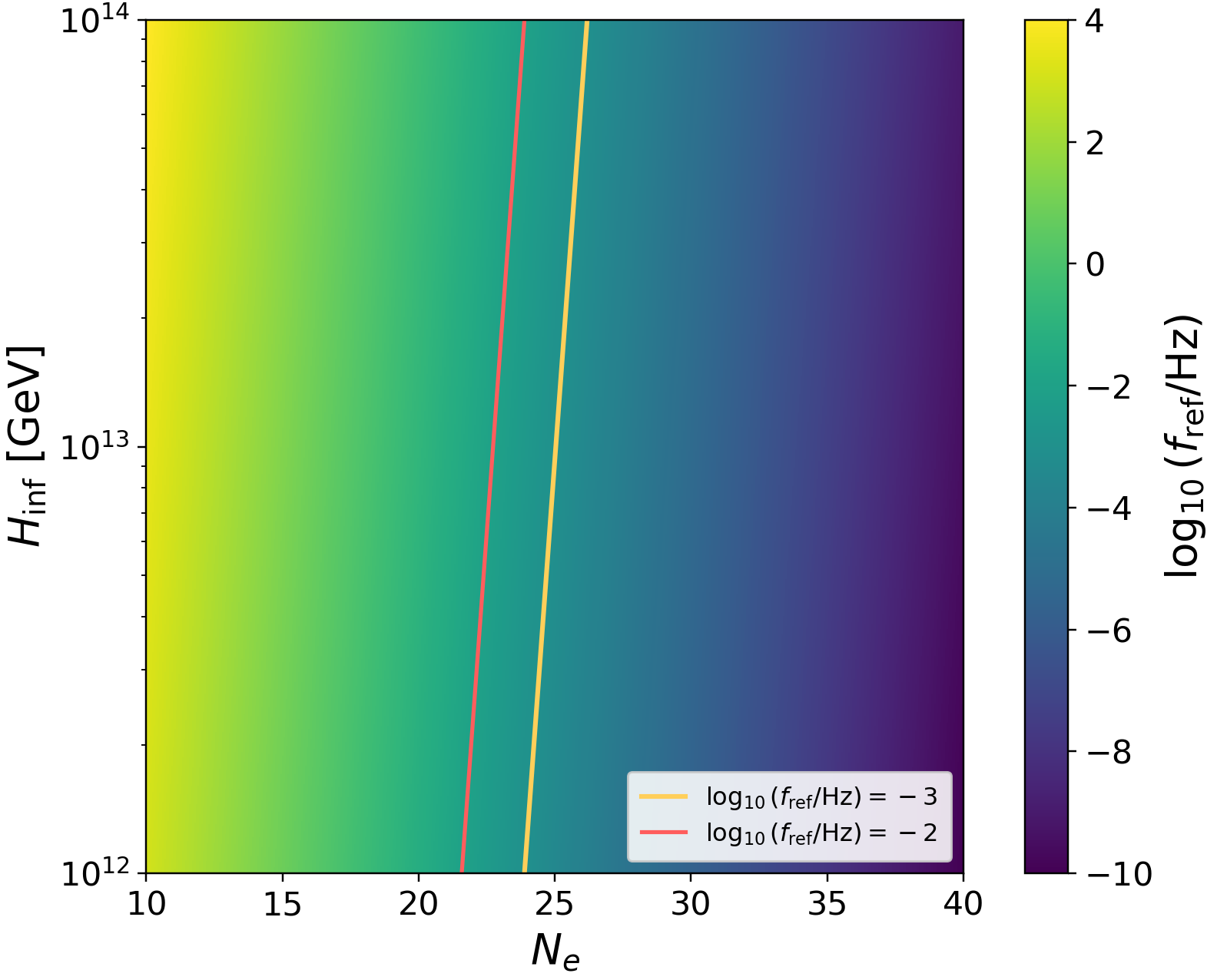}
    \vspace{-0.5em}
    \caption{
    Mapping between the InPT epoch $N_e$, the inflationary scale $H_{\rm inf}$, and the present-day reference frequency $f_{\rm ref}$. The point $f_{\rm ref}=10^{-3}\,{\rm Hz}$ and $H_{\rm inf}=10^{14}\,{\rm GeV}$ corresponds to $N_e\simeq26.18$. The gray dashed horizontal line marks the Planck+BK15 benchmark value $H_{\rm inf}\simeq5.9\times10^{13}\,{\rm GeV}$.
    }
    \label{fig:ne_hinf_scan}
    \vspace{-1.2em}
\end{figure}

For the representative value $f_{\rm ref}=10^{-3}\,{\rm Hz}$ and $H_{\rm inf}=10^{14}\,{\rm GeV}$, this gives $N_e=40-\ln(10^6)\simeq26.18$. Thus, the statement that Taiji-band InPT signals correspond to transitions about 26 e-folds before the end of inflation assumes the benchmark inflation scale $H_{\rm inf}=10^{14}\,{\rm GeV}$. Figure~\ref{fig:ne_hinf_scan} illustrates how the frequency--epoch mapping shifts with $H_{\rm inf}$. Because $N_e$ depends logarithmically on $H_{\rm inf}$, even if $H_{\rm inf}$ changes by orders of magnitude, $N_e$ remains around 25.

Finally, we comment on computational cost. The NS runs were performed with \texttt{Bilby.dynesty}. For the formal InPT runs with ten free parameters, we used $n_{\rm live}=1000$, \texttt{walks}=64, $\Delta\log Z=0.1$, and one sampler process. The likelihood evaluation used GPU acceleration for the frequency-domain signal and noise calculations, with $N_0=126$ independent data segments of duration $T=10^6\,{\rm s}$ and $5\times10^5$ frequency samples over $3\times10^{-5}\leq f\leq0.5\,{\rm Hz}$. With this GPU-accelerated implementation, the Table~\ref{ebl} runs that include the InPT signal and the background components, with ten free parameters in total, required about 9--36 h of wall time, while the corresponding evidence calculations for the model without an InPT signal typically required about 1--3 h. This computational cost is why we use NS for representative injections and FIM forecasts for dense contour maps.

\FloatBarrier

\section{Conclusion}\label{qmg}

In this work, we investigated the detectability and parameter reconstruction of SGWBs generated by InPTs using a Taiji-like space-based detector. Focusing on the secondary GW component, we adopted a minimal, model-independent parametrization in terms of $(B_{\rm ref}, f_{\rm ref})$ and constructed a realistic data-analysis framework that incorporates instrumental noise, astrophysical foregrounds and backgrounds, and the full TDI responses in $A$, $E$, and $T$ channels. Using both Fisher forecasts and full Bayesian inference with NS, we quantified parameter uncertainties and evaluated the Bayesian evidence for signal detection.

Our results demonstrate a clear separation among the exclusion, detection, and reliable parameter-recovery regimes. While signals with $\mathrm{SNR}\gtrsim10$ can satisfy conventional detection criteria, robust reconstruction of the InPT spectral parameters requires significantly stronger signals, corresponding to $\ln \mathrm{BF}\sim \mathcal{O}(10)$ for the scenarios considered. The added full-plane FIM maps show where detected signals can be spectrally characterized. We also showed how $B_{\rm ref}$ and $f_{\rm ref}$ can be mapped to physical quantities only after specifying assumptions about $\epsilon$, $L/\rho_{\rm inf}$, $\phi_0$, and $H_{\rm inf}$. Supplementary checks with alternative representative spectral shapes support the same reconstruction picture, while showing that the detailed relation between SNR and Bayesian evidence can depend on the template shape. We further show that astrophysical foregrounds degrade parameter precision and shift the boundaries between these regimes. Overall, our analysis provides a quantitative benchmark for assessing the capability of future space-based missions to probe InPTs through stochastic gravitational radiation.

\FloatBarrier

\section*{Acknowledgments}
We thank Ju Chen, Ming-Hui Du, and Chang Liu for helpful discussions. H.~A. is supported by the National Key R\&D Program of China under Grants No.~2021YFC2203100 and No.~2017YFA0402204, the National Science Foundation of China (NSFC) under Grants No.~12475107 and No.~12525506, and the Tsinghua University Dushi Program. H.-K.~G. is supported by the startup fund provided by the University of Chinese Academy of Sciences and by the NSFC under Grants No.~12547104 and No.~12475109.

\section*{Data Availability}
The data that support the findings of this article are not publicly available upon publication because it is not technically feasible and/or the cost of preparing, depositing, and hosting the data would be prohibitive within the terms of this research project. The data are available from the authors upon reasonable request.

\appendix

\section{Supplementary Parameter-Estimation Checks}\label{app:supplementary_pe}

This appendix presents two additional PE checks for the reconstruction results. These checks are not used to define the exclusion, detection, or recovery contours in the main text; instead, they show how the reconstruction behaves in cases slightly different from the main benchmark. First, we lower the InPT amplitude to $\log_{10}B_{\rm ref}=-14.4$ at $\log_{10}(f_{\rm ref}/{\rm Hz})=-3$. This signal is closer to the recovery threshold, so it provides a direct test of whether the FIM still gives the right uncertainty scale when the full PE posterior becomes broader and less Gaussian.

\begin{figure}[H]
    \centering
    \setlength{\fboxsep}{0pt}
    \includegraphics[width=0.63\linewidth]{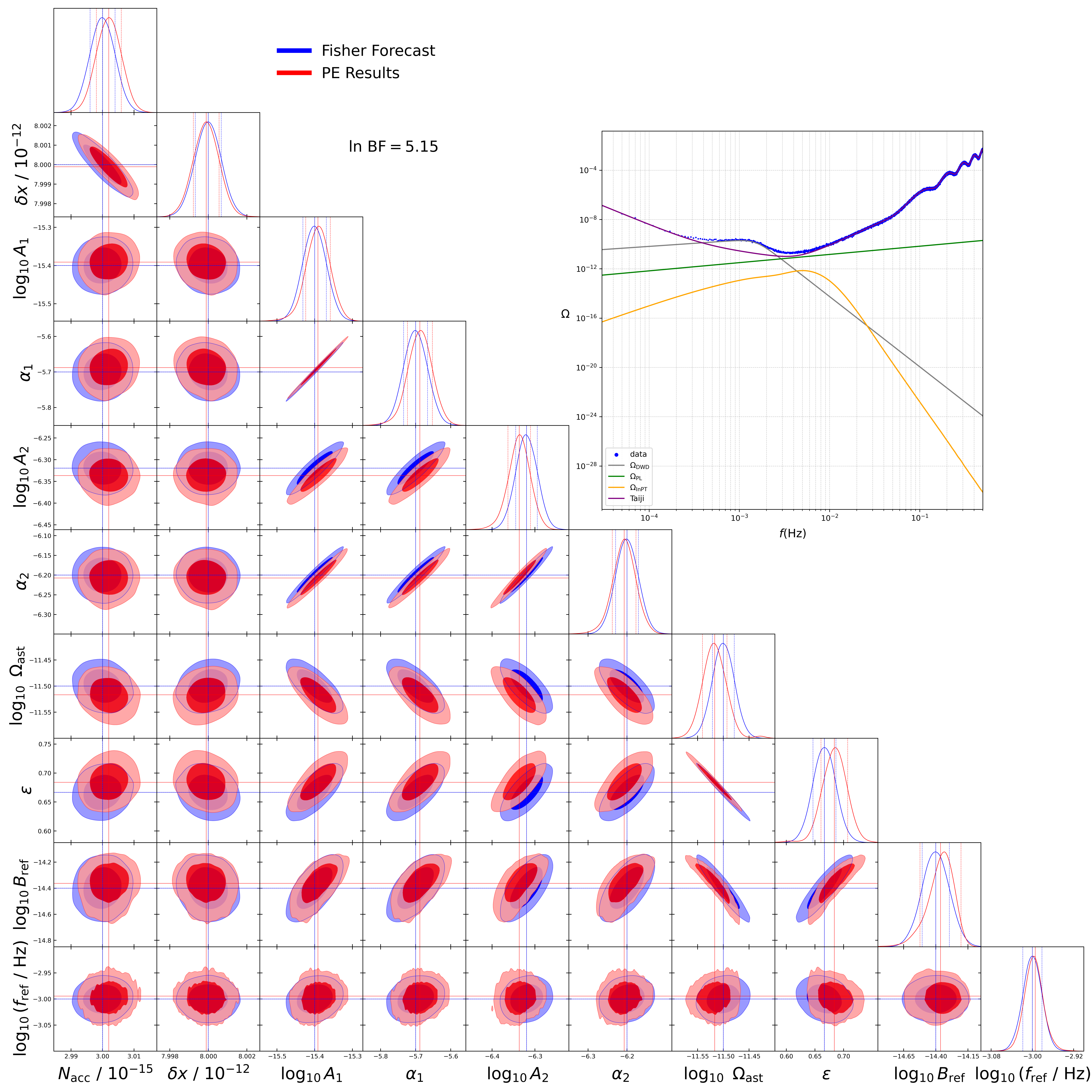}
    \caption{
    Lower-SNR comparison between the PE posterior and the FIM forecast at $\log_{10}B_{\rm ref}=-14.4$ and $\log_{10}(f_{\rm ref}/{\rm Hz})=-3$. This injection has $\mathrm{SNR}_a=46.860$, $\mathrm{SNR}_r=26.354$, and $\ln\mathrm{BF}=5.155$. Compared with the high-SNR example in Fig.~\ref{egl}, the PE posterior is broader and more visibly non-Gaussian, while the FIM still captures the leading uncertainty scale.
    }
    \label{fig:lower_snr_pe}
\end{figure}

\begin{table}[H]
\centering
\begin{tabular}{lcccc}
\hline\hline
Parameter & Fiducial value & PE recovery & Fisher unc.\ (\%) & PE unc.\ (\%) \\
\hline
$N_{\rm acc} / 10^{-15}$
& $3.000$ & $3.002$ & $0.132$ & $0.132$ \\
$\delta x / 10^{-12}$
& $8.000$ & $8.000$ & $0.008$ & $0.008$ \\
$\log_{10} A_1$
& $-15.400$ & $-15.391$ & $0.202$ & $0.212$ \\
$\alpha_1$
& $-5.700$ & $-5.688$ & $0.598$ & $0.624$ \\
$\log_{10} A_2$
& $-6.320$ & $-6.337$ & $0.397$ & $0.415$ \\
$\alpha_2$
& $-6.200$ & $-6.207$ & $0.471$ & $0.490$ \\
$\log_{10} \Omega_{\rm ast}$
& $-11.500$ & $-11.517$ & $0.185$ & $0.207$ \\
$\varepsilon$
& $0.667$ & $0.684$ & $3.017$ & $3.401$ \\
$\log_{10} B_{\rm ref}$
& $-14.400$ & $-14.363$ & $0.733$ & $1.112$ \\
$\log_{10} (f_{\rm ref} / \rm Hz)$
& $-3.000$ & $-2.994$ & $0.622$ & $4.539$ \\
\hline\hline
\end{tabular}
\caption{Injected parameter values, posterior means inferred from the lower-SNR PE run at $\log_{10}B_{\rm ref}=-14.4$, and relative uncertainties obtained from the FIM and PE analyses. For this injection, $\mathrm{SNR}_a=46.860$, $\mathrm{SNR}_r=26.354$, and $\ln\mathrm{BF}=5.155$.}
\label{tab:pe144_fim_pe}
\end{table}

Figure~\ref{fig:lower_snr_pe} and Table~\ref{tab:pe144_fim_pe} show that near the recovery boundary the signal remains detectable while the transition frequency is less sharply localized; the FIM should, therefore, be read as an uncertainty-scale forecast rather than a full description of the posterior shape.

\clearpage

\begin{figure}[H]
	    \centering
	    \setlength{\fboxsep}{0pt}
	    \makebox[\linewidth][c]{%
	    \includegraphics[width=0.47\linewidth]{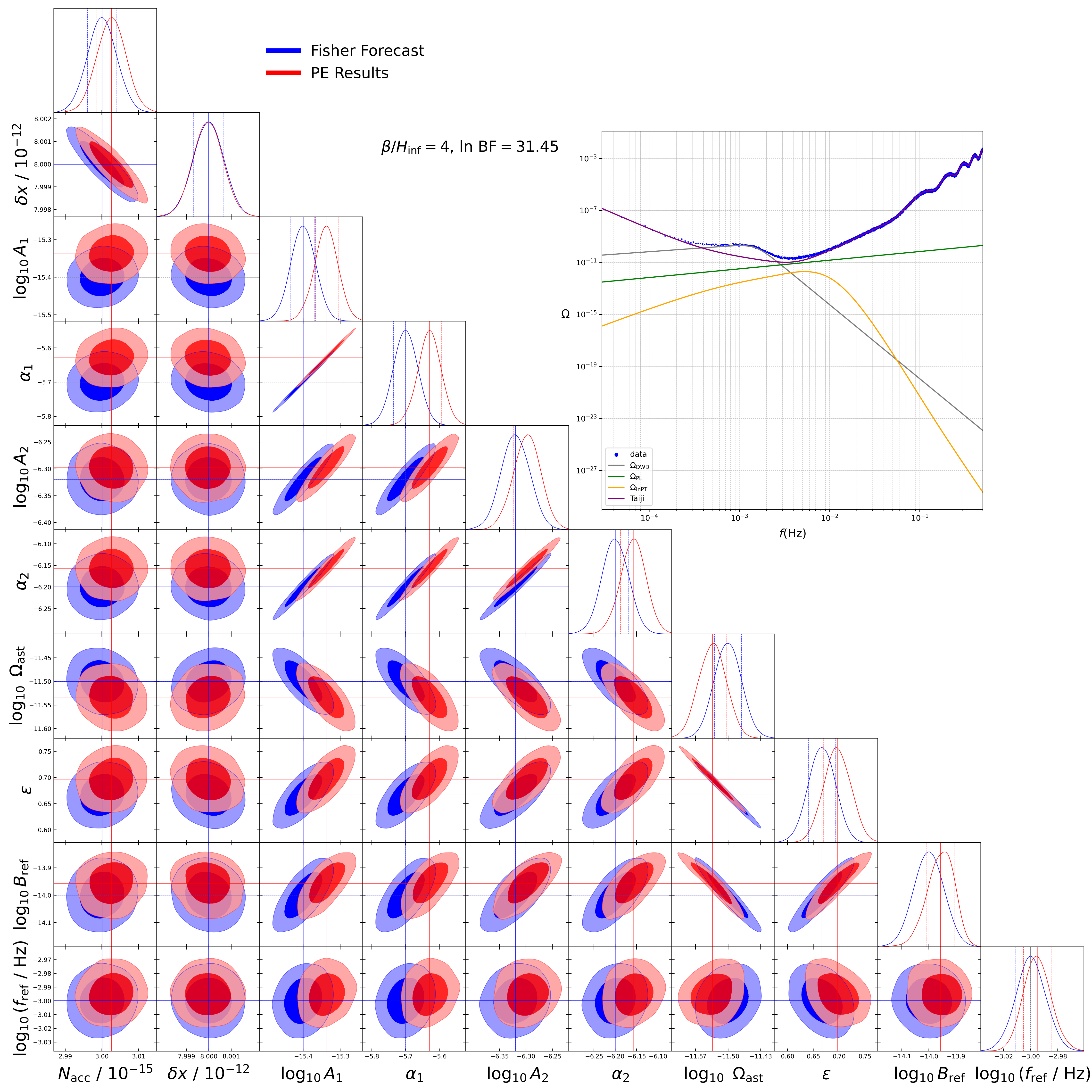}
	    \hspace{0.006\linewidth}
	    \includegraphics[width=0.47\linewidth]{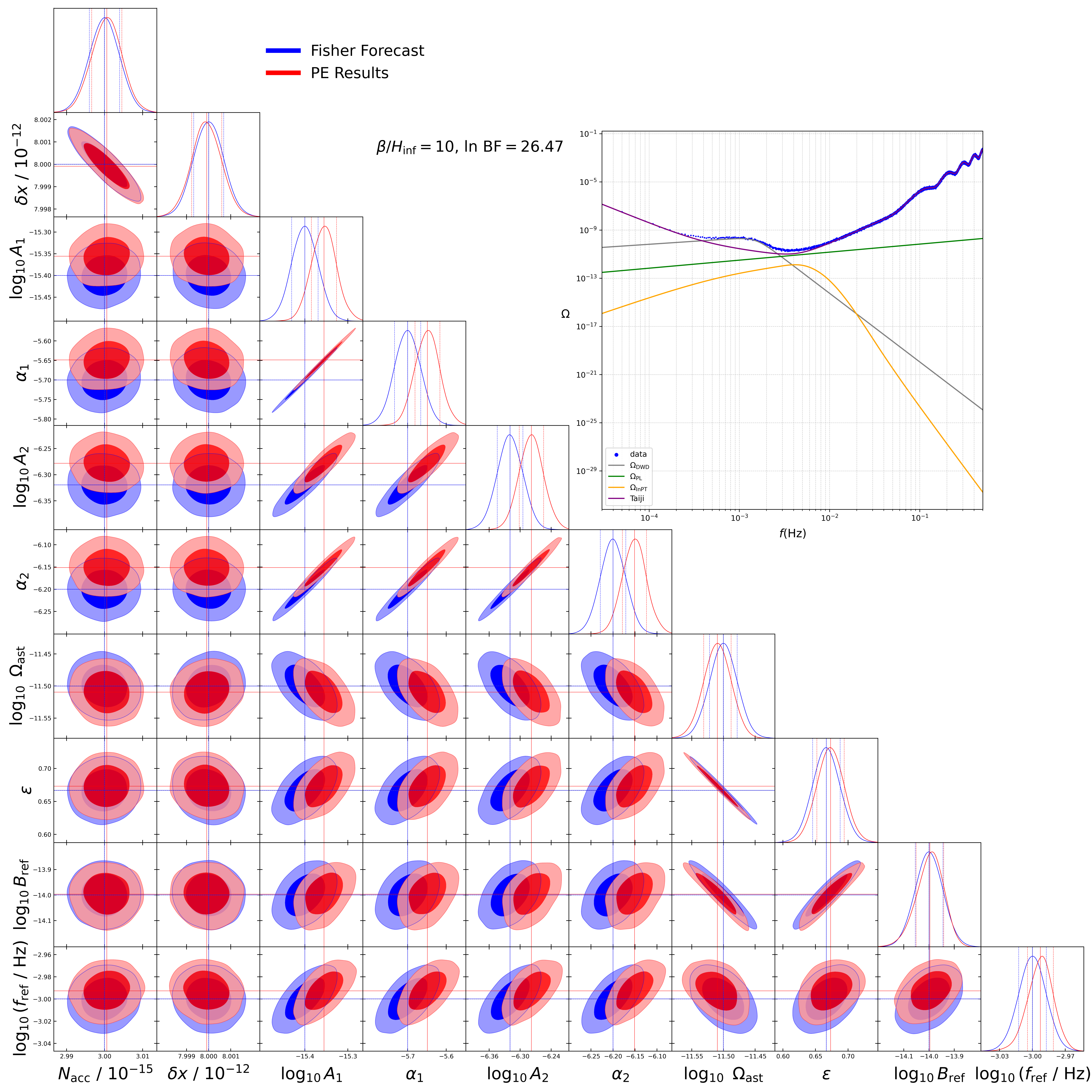}}
    \caption{
    Representative PE-vs-FIM checks for the fixed-shape benchmark extended to $\beta/H_{\rm inf}=4$ (left) and $\beta/H_{\rm inf}=10$ (right), both at the fiducial injection $(\log_{10}B_{\rm ref},\log_{10}(f_{\rm ref}/{\rm Hz}))=(-14,-3)$. The final PE--BF reruns give $\ln\mathrm{BF}=31.454$ for $\beta/H_{\rm inf}=4$ and $\ln\mathrm{BF}=26.467$ for $\beta/H_{\rm inf}=10$. In both cases, the PE posteriors remain close to the injected point and agree well with the corresponding FIM uncertainty scale, illustrating that these alternative templates preserve the fixed-shape recovery picture while changing the detailed spectral shape and evidence.
    }
    \label{fig:beta_pe_checks}
\end{figure}

\enlargethispage{3.5\baselineskip}

\begin{table}[H]
\centering
\small
\setlength{\tabcolsep}{4pt}
\renewcommand{\arraystretch}{0.92}
\resizebox{\linewidth}{!}{%
\begin{tabular}{lccccccc}
\hline\hline
\multirow{2}{*}{Parameter} &
\multirow{2}{*}{Fiducial} &
\multicolumn{3}{c}{$\beta/H_{\rm inf}=4$} &
\multicolumn{3}{c}{$\beta/H_{\rm inf}=10$} \\
\cline{3-8}
& & PE recovery & Fisher unc.\ (\%) & PE unc.\ (\%) & PE recovery & Fisher unc.\ (\%) & PE unc.\ (\%) \\
\hline
$N_{\rm acc} / 10^{-15}$ & $3.000$ & $3.003$ & $0.133$ & $0.132$ & $3.001$ & $0.132$ & $0.133$ \\
$\delta x / 10^{-12}$ & $8.000$ & $8.000$ & $0.008$ & $0.008$ & $8.000$ & $0.008$ & $0.008$ \\
$\log_{10} A_1$ & $-15.400$ & $-15.338$ & $0.218$ & $0.210$ & $-15.356$ & $0.200$ & $0.191$ \\
$\alpha_1$ & $-5.700$ & $-5.629$ & $0.639$ & $0.621$ & $-5.649$ & $0.593$ & $0.571$ \\
$\log_{10} A_2$ & $-6.320$ & $-6.298$ & $0.430$ & $0.413$ & $-6.278$ & $0.390$ & $0.376$ \\
$\alpha_2$ & $-6.200$ & $-6.158$ & $0.507$ & $0.487$ & $-6.151$ & $0.465$ & $0.447$ \\
$\log_{10}\Omega_{\rm ast}$ & $-11.500$ & $-11.533$ & $0.252$ & $0.256$ & $-11.510$ & $0.187$ & $0.188$ \\
$\varepsilon$ & $0.667$ & $0.697$ & $3.907$ & $3.789$ & $0.673$ & $3.126$ & $3.108$ \\
$\log_{10} B_{\rm ref}$ & $-14.000$ & $-13.957$ & $0.397$ & $0.363$ & $-13.996$ & $0.386$ & $0.390$ \\
$\log_{10}(f_{\rm ref}/{\rm Hz})$ & $-3.000$ & $-2.995$ & $0.368$ & $0.342$ & $-2.993$ & $0.420$ & $0.395$ \\
\hline\hline
\end{tabular}
}
\caption{All-parameter PE-vs-FIM comparison for the two corner plots in Fig.~\ref{fig:beta_pe_checks}. The left set of columns gives the results for the fixed-shape template with $\beta/H_{\rm inf}=4$, and the right set gives the results for $\beta/H_{\rm inf}=10$. For $\beta/H_{\rm inf}=4$, $\mathrm{SNR}_a=135.293$, $\mathrm{SNR}_r=76.232$, and $\ln\mathrm{BF}=31.454$; for $\beta/H_{\rm inf}=10$, $\mathrm{SNR}_a=95.188$, $\mathrm{SNR}_r=50.122$, and $\ln\mathrm{BF}=26.467$.}
\label{tab:beta_allparams}
\end{table}

\begin{table}[H]
\centering

\small
\setlength{\tabcolsep}{3pt}
\renewcommand{\arraystretch}{0.92}
\begin{tabular*}{\linewidth}{@{\extracolsep{\fill}}lccccccc@{}}
\hline\hline
\multirow{2}{*}{$\beta/H_{\rm inf}$} &
\multirow{2}{*}{$\mathrm{SNR}_a$} &
\multirow{2}{*}{$\mathrm{SNR}_r$} &
\multirow{2}{*}{$\ln\mathrm{BF}$} &
\multicolumn{2}{c}{$\log_{10}B_{\rm ref}$} &
\multicolumn{2}{c}{$\log_{10}(f_{\rm ref}/{\rm Hz})$} \\
\cline{5-8}
& & & & PE recovery & PE unc.\ (\%) & PE recovery & PE unc.\ (\%) \\
\hline
$4$ & $135.293$ & $76.232$ & $31.454$ & $-13.957$ & $0.363\%$ & $-2.995$ & $0.342\%$ \\
$5$ & $117.707$ & $66.197$ & $42.236$ & $-14.014$ & $0.312\%$ & $-3.001$ & $0.256\%$ \\
$10$ & $95.188$ & $50.122$ & $26.467$ & $-13.996$ & $0.390\%$ & $-2.993$ & $0.395\%$ \\
\hline\hline
\end{tabular*}
\caption{Compact summary of representative parameter-estimation checks for fixed-shape templates. The SNR definitions are given in Sec.~\ref{ohb}, and the all-parameter PE-vs-FIM comparison is given in Table~\ref{tab:beta_allparams}. Unlike a single-template amplitude scan, these rows keep the same fiducial $(B_{\rm ref},f_{\rm ref})$ and change only the template shape through $\beta/H_{\rm inf}$; therefore, $\ln\mathrm{BF}$ need not follow the SNR ordering.}
\label{tab:beta_pe_checks}

\end{table}

Figure~\ref{fig:beta_pe_checks} examines the fixed-shape assumption used in the main inference. The three templates use the same fiducial $(B_{\rm ref},f_{\rm ref})$, so the comparison tests spectral-shape dependence rather than a change in peak amplitude. Tables~\ref{tab:beta_allparams} and \ref{tab:beta_pe_checks} show that the posterior means remain close to the injected values and that the PE uncertainties agree with the FIM forecasts. Because the shape changes from row to row, the Bayes factor need not follow the usual amplitude-scan SNR ordering.

\FloatBarrier

\section{Validity of the Secondary-Only Approximation}\label{app:primary_secondary}

\begin{figure}[H]
    \centering
    \setlength{\fboxsep}{0pt}
    \includegraphics[width=0.62\linewidth]{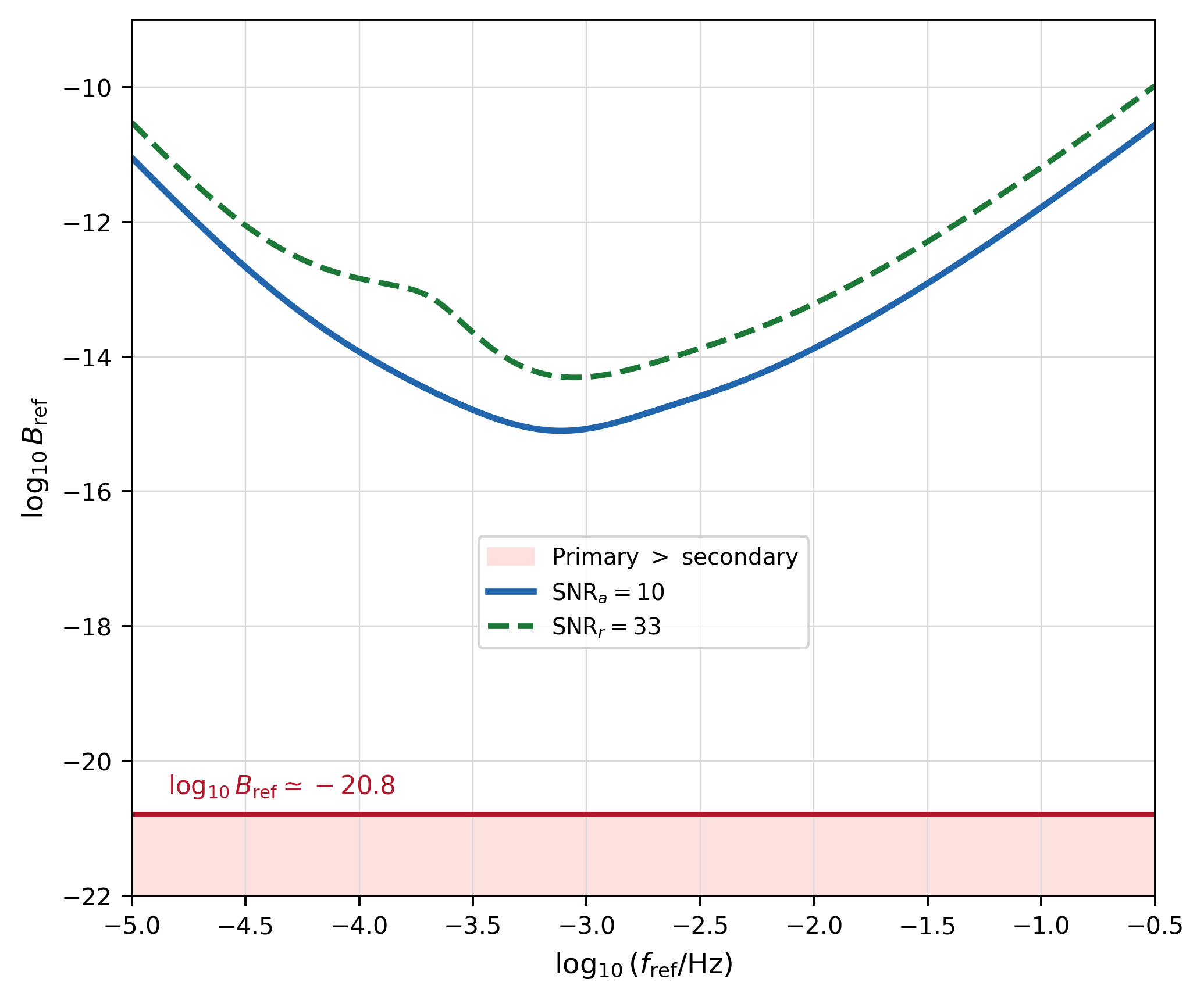}
    \caption{
    Validity range of the secondary-only approximation. The red boundary marks the approximate region where the primary and secondary peak amplitudes become comparable for the fiducial choices stated in the text. The boundary lies near $\log_{10}B_{\rm ref}\simeq-20.8$, many orders of magnitude below the Taiji-sensitive region shown by the SNR contours in Fig.~\ref{elv}.
    }
    \label{fig:primary_secondary_boundary}
\end{figure}

\begingroup
\setlength{\abovedisplayskip}{5pt plus 2pt minus 2pt}
\setlength{\belowdisplayskip}{5pt plus 2pt minus 2pt}
\setlength{\abovedisplayshortskip}{4pt plus 2pt minus 2pt}
\setlength{\belowdisplayshortskip}{4pt plus 2pt minus 2pt}

Here, we collect the quantitative check of the secondary-only approximation. The comparison is placed after the Taiji detectability criteria because Fig.~\ref{fig:primary_secondary_boundary} is meant to be read against the SNR contours in Fig.~\ref{elv}. Defining $R_{\rm pri/sec}\equiv\Omega_{\rm pri}^{\rm peak}/\Omega_{\rm sec}^{\rm peak}$ and using the rounded peak scalings in Eqs.~\eqref{eq:omega_pri_peak} and \eqref{eq:omega_sec_peak}, we find
\begin{equation}
R_{\rm pri/sec}
\simeq
7.0\times10^{-7}
\epsilon^2
\left(\frac{\phi_0}{M_{\rm Pl}}\right)^4
\left(\frac{\beta}{H_{\rm inf}}\right)
\left(\frac{\rho_{\rm inf}}{L}\right)^2 .
\end{equation}
The primary component would dominate only for
\begin{equation}
\frac{L}{\rho_{\rm inf}}
\lesssim
8.4\times10^{-4}\,
\epsilon
\left(\frac{\phi_0}{M_{\rm Pl}}\right)^2
\left(\frac{\beta}{H_{\rm inf}}\right)^{1/2}.
\end{equation}
For the fiducial example $\epsilon=10^{-2}$, $\phi_0=M_{\rm Pl}$, and $\beta/H_{\rm inf}=5$, this gives $L/\rho_{\rm inf}\simeq1.9\times10^{-5}$, or $\log_{10}B_{\rm ref}\simeq -20.8$ after using $B_{\rm ref}=\Omega_{\rm R}A_{\rm ref}^2$. Figure~\ref{fig:primary_secondary_boundary} shows that this boundary is far below the Taiji-sensitive region in Fig.~\ref{elv}. The secondary-only approximation is, therefore, valid throughout the Taiji-accessible region probed in this work.

This large separation indicates that the parameter region where the primary contribution could become comparable to the secondary one is outside the Taiji-accessible range considered here. Hence, within the detectable and reconstructable region of this study, the InPT signal is safely described by the secondary GW component.

\endgroup

\FloatBarrier

\bibliographystyle{JHEP}
\bibliography{refer}

\end{document}